\documentclass[fullpage]{article}       
\usepackage{amssymb,amsmath}
\usepackage{graphicx}
\usepackage{todonotes}
\usepackage{hyperref}
\usepackage{array}
\usepackage{multirow}
\usepackage{xspace}
\usepackage{booktabs}
\usepackage{siunitx}
\usepackage{subcaption}
\usepackage{caption}
\usepackage[numbers]{natbib}
\usepackage{bold-extra}

\usepackage{longtable,tabulary}
\newcommand\TODOGRETA[1]{\textcolor{black}{#1}}
\newcommand\grtext[1]{{\color{black}{#1}}}

\newcommand\rtext[1]{{\color{black}{#1}}}

\newcommand\ele[1]{\texttt{#1}}

\newcommand{\rqone}{\textbf{RQ1}\xspace}
\newcommand{\rqtwo}{\textbf{RQ2}\xspace}
\newcommand{\rqthree}{\textbf{RQ3}\xspace}
\newcommand{\rqfour}{\textbf{RQ4}\xspace}
\newcommand{\rqfive}{\textbf{RQ5}\xspace}

\newcommand{\psnumber}{36\xspace}
\newcommand{\gen}{\textbf{Ind}\xspace}
\newcommand{\lsp}{\textbf{Dep}\xspace}
\newcommand{\bpml}{\textbf{BPM}\xspace}
\newcommand{\nobpml}{\textbf{NoBPM}\xspace}
\newcommand{\oldl}{\textbf{Exist}\xspace}
\newcommand{\newl}{\textbf{New}\xspace}
\newcommand{\frm}{\textbf{FRM}\xspace}

\newcommand{\exe}{\textbf{Exe}\xspace}
\newcommand{\model}{\textbf{Mod}\xspace}
\newcommand{\modexe}{\textbf{ModExe}\xspace}

\newcommand{\dm}{\textbf{Dom}\xspace}

\newcommand{\eval}{\textbf{Eval}\xspace}
\newcommand{\proc}{\textbf{Proc}\xspace}
\newcommand{\dec}{\textbf{Dec}\xspace}
\newcommand{\act}{\textbf{Act}\xspace}
\newcommand{\art}{\textbf{Art}\xspace}
\begin{document}

\title{What's My Process Model Composed of?\\
A Systematic Literature Review of Meta-Models in BPM
}


\author{Greta Adamo$^{1,2}$ \and Chiara Ghidini \and Chiara Di Francescomarino\\
$^1$ Fondazione Bruno Kessler, Trento, Italy.\\
$^2$ University of Genoa, Genoa, Italy.}


\date{}

\maketitle

\begin{abstract}
Business process modelling languages typically enable the representation of business process models by employing (graphical) symbols. These symbols can vary depending upon the verbosity of the language, the modeling paradigm, the focus of the language, and so on. 
To make explicit the different constructs and rules employed by a specific language as well as bridge the gap across different languages, meta-models have been proposed in literature. These meta-models are a crucial source of knowledge on what state-of-the-art literature considers relevant to describe business processes.
Moreover, the rapid growth of techniques and tools that aim at supporting all dimensions of business processes and not only its control flow perspective, as for instance data and organisational aspects, makes even more important to have a clear idea, already at the conceptual level, of the key process constructs.
The goal of this work is to provide the first extensive \textit{systematic literature review} (SLR) of business process meta-models. This SLR aims at answering research questions concerning: (i) the kind of meta-models proposed in literature; (ii) the recurring constructs they contain; (iii) their purposes; and (iv) their evaluations.
The SRL was performed manually considering papers automatically retrieved from reference paper repositories as well as proceedings of the main conferences in the Business Process Management research area. 
Thirty-six papers were selected and evaluated against four research questions. The results indicate the existence of a reasonable body of work conducted in this specific area, but not a full maturity. In particular, while traditional paradigms towards business process modelling, and aspects related to the business process control flow seem to be well present, novel paradigms and aspects related to the organisational, data and goal-oriented aspects of business processes seem to be still under-investigated.
\end{abstract}

\section{Introduction}
\label{introduction}

Business process modelling languages (BPMLs) typically enable the representation of business processes via the creation of process models, 
 which are constructed using the elements and graphical symbols of the BPML itself. A process \emph{model} is a conceptual$\backslash$abstract representation of a business process, whose goal is to describe or prescribe a \textit{real} process by specifying how the process should/could/might be performed. 
The different constructs and rules employed by a specific BPML to create models are contained in the business process meta-model (BPMM). By quoting \citet[pg.~76]{DBLP:books/daglib/0029914}
\begin{quote}
``Models are expressed in metamodels that are associated with notations, often of graphical nature. For instance the Petri net metamodel consists of places and transitions that form a directed bipartite graph. The traditional Petri net notation associates graphical symbols with metamodel elements. For instance, places are represented by circles, transitions by rectangles, and the graph structure by directed edges. ''
\end{quote}
Meta-models can be therefore conceived as a descriptive systematisation of abstract categories of the \textit{world} (the business process in our case).

Due to the number of BPMLs available in literature, a number of associated meta-models exist. These meta-models can vary greatly, reflecting the expressive power of the language, its specificities in terms of the specific sub-domain it may focus at, or the modelling paradigm the BPML adheres to. 
Meta-models are also defined in literature independently from specific BPMLs with the aim of ``navigating'' across the different BPMLs, bridge the gap across them, foster a common ground across different notations, and promote interoperability, thus further increasing their overall number.   
Besides the specific purposes for which they are introduced, meta-models are a crucial source of knowledge on the constructs and rules that state-of-the-art literature considers relevant to model (and thus describe) business processes; yet a detailed analysis of business process meta-models described in literature is still absent. 

Moreover, the growth of approaches and tools aiming at supporting business processes in a multi-perspective manner by looking beyond  the control-flow perspective and including other dimensions, such as the data, organisational and goal oriented ones, shows that the time is now ripe to focus on an investigation of different types of process constructs also at the conceptual level. Indeed, even though all the most popular definitions of business process contain aspects that go beyond the control flow, some of them are still neglected, or not clearly described, in state-of-the-art meta-models.

The goal of this work is to provide the first comprehensive \textit{Systematic Literature Review (SLR)} of business process modelling language meta-models in the BPM field. The SLR aims at identifying, categorising, and describing works related to business process meta-models. It focuses on works in the BPM research area and has been driven by four research questions concerning (i) the kind of meta-models proposed in literature; (ii) the recurring constructs they contain; (iii) the purpose(s) of the proposed meta-models; and finally (iv) their evaluations.
In addition to providing a first comprehensive analysis of business process modelling language meta-models, the SLR provides a first example of a framework to categorise, compare and analyse BPML meta-models.  

The paper is organised as follow: in Section~\ref{sec:method} the method employed to perform the SLR is presented, by describing both the planning and the conducting  of the review. Special emphasis is given in this section to the definition of the research questions (Section~\ref{sec:RQ}), and to the protocol of review with the description of inclusion and exclusion criteria (Section~\ref{sec:protocol}). The results of our data (papers) collection and selection are given in Section~\ref{sec:results}, and a brief summary of the \psnumber selected papers we retained for answering to the research questions is given in Section~\ref{sec:summaryofpapers}. A detailed answer to the four research questions is provided in Section~\ref{sec:RRQs}, followed by an extensive discussion in Section \ref{sec:discussion}. Final remarks and conclusions are presented in Section~\ref{sec:conclusion}.

\section{Method}
\label{sec:method}


\begin{figure}[t]
\centering
\includegraphics[width=.80\textwidth]{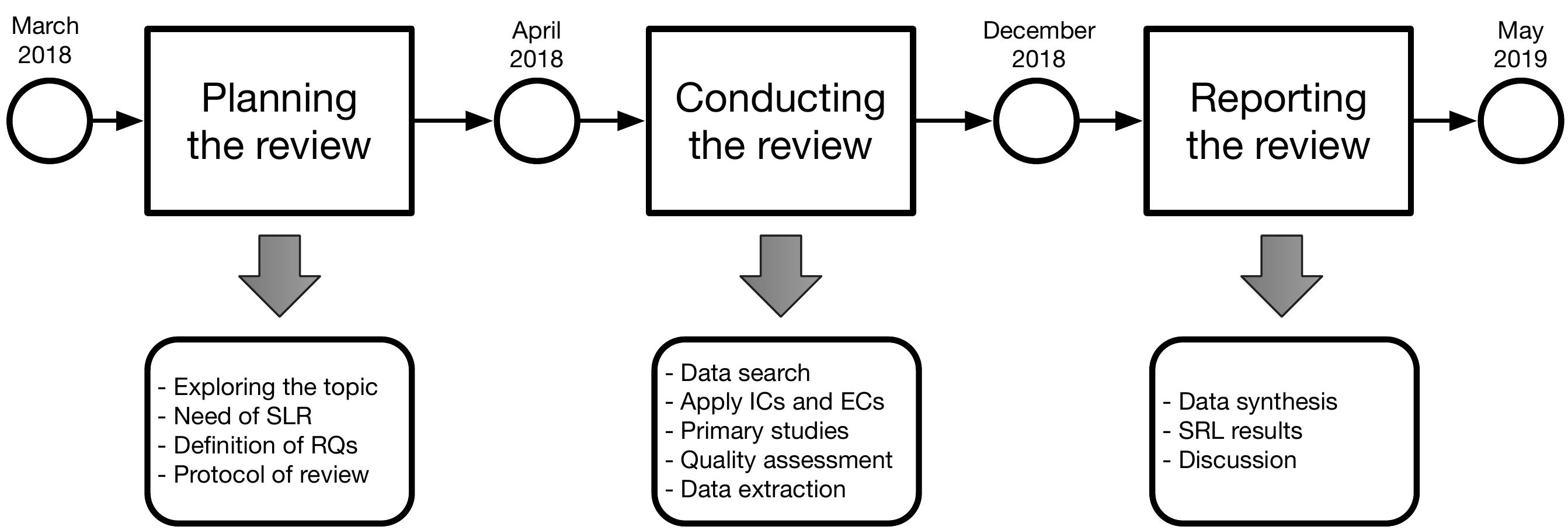}
\caption{Method used for the SLR}
\label{fig:method}
\end{figure}

The SLR presented in this paper follows the guidelines for conducting a SLR proposed in~\cite{kitchenham2004procedures, Kitchenham07guidelinesfor}. Following these guidelines, a SLR is divided in three pivotal phases, graphically summarised in Figure~\ref{fig:method}: \emph{planning} the review; \emph{conducting} the review; and \emph{reporting}. 

The remaining of this section is devoted to the description of the \emph{planning} of this SLR with particular emphasis to: a brief description of the aspects of the topic relevant to the SLR, the motivations for performing it, the development of the research questions, and the definition of the specific protocol for review adopted. The \emph{conduction} of the review, and its results are described in the next three sections, while this paper, and the additional material linked in the paper, constitutes the SLR \emph{report}. 

\subsection{Background}
\label{sec:background}

The most modern and popular definition of business process is likely the one provided by \citet{DBLP:books/daglib/0029914}:
\begin{quote}
	\label{def:process}
	``a set of activities that are performed in coordination in an organizational and technical environment. These activities jointly realize a business goal. Each business process is enacted by a single organization, but it may interact with business processes performed by other organizations.''
\end{quote}
Thus, business processes are composed of a set of interacting participants (activities, organisational roles, goals just to mention a few\footnote{A comparative analysis of elements belonging to different business process modelling notations can be found in \cite{Adamo:2017:AIIA}}) which are typically captured by business process models specified in a business process modelling languages. These languages, and the underlying modelling methodologies, often steam from the conceptual modelling field, whose aim is to identify, analyse, capture, and describe the basic concepts and features of a domain (\textit{universe of discourse}) \cite{32d47c9b496748ff911fb88726992462}. 

Within the last 15 years, an increasing effort has been spent
 in providing business process modelling techniques, methodologies, as well as tools and languages for the representation of business process models \cite{DBLP:journals/csur/MiliTJLEE10,AGUILARSAVEN2004129}. Focusing on the modelling languages, it is easy to observe their great variability in terms of constructs and rules they offer to compose process models. These differences have different causes ranging from the different expressive powers of the languages, to the specific sub-domain it may focus at, or even the specific modelling paradigm and approach the modelling language adheres to. Think for instance at the different constructs and rules offered by languages such as BPMN 2.0\footnote{\url{https://www.omg.org/spec/BPMN/2.0/About-BPMN/}}, UML-AD\footnote{\url{https://www.omg.org/spec/UML/About-UML/}}, EPCs~\cite{scheer2013aris}, DECLARE~\cite{Pesic:2007:DFS:1317532.1318056}, YAWL~\cite{YAWL:2010aa}, and CMMN\footnote{\url{https://www.omg.org/spec/CMMN/About-CMMN/}}, partly due to their declarative vs imperative, activity- vs artefact-centric nature.
 We introduce here some modelling paradigms that will be mentioned throughout the SLR. 

\paragraph{Imperative and Declarative Paradigms}
Imperative BPMLs enable designing process models by specifying the allowed flows of activities.
 Thus, imperative languages such as BPMN, EPC, UML-AD, and YAWL provide particular elements to denote the start and the end of a process \cite{Adamo:2017:AIIA} and  force the production of process models that specify all the possible ways the control flow moves from the start towards the end element. It has been shown that this kind of paradigm is suitable for predictable processes with few variations, but that it is not so effective in situations in which there are many variabilities \cite{DBLP:conf/caise/GiacomoDMM15}.




Declarative BPMLs, such as DECLARE and CMMN, have hence been proposed. As reported in~\cite{DBLP:conf/caise/GiacomoDMM15}, these languages allow modellers to (only) specify constraints on the allowed flows, that is, unless a flow does not satisfy the provided constraints, it is allowed. 
 As a consequence, declarative languages, such as DECLARE, focus on how to express relationships (constraints) between specific process participants rather than modelling a comprehensive view of the control flow with e.g. a well defined start and a well defined end\footnote{As an example DECLARE provides constructs for a first / last  activity, but it does not force neither suggest a process model should always include
 them.} \cite{DBLP:conf/caise/GiacomoDMM15}. 
Traditional BPMLs follow the imperative paradigm.

\paragraph{Activity-centric and Artefact-centric Paradigms}
Activity-centric BPMLs see the process control flow as a series of activities that enable the process to move from the start towards the end construct.
 Thus, as reported in~\cite{DBLP:conf/bpm/0001W13}, languages such as BPMN, UML-AD, YAWL, and DECLARE use activities and control structures (i.e., gateways) as
primary modelling elements, while considering data objects as secondary
 components, often used as pre- and post-conditions for the execution of an activity, or as decision indicators in case of control structure conditions~\cite{DBLP:conf/bpm/0001W13}.

 Differently, artefact-centric process modelling, as for instance CMMN, considers data objects and their life-cycles as primary
 modelling elements, and activities are of importance as they participate to an object change of state. This, more recent, paradigm has been developed and proved useful in scenarios where the flow of the process is originated from the data objects, as for example in case of 
 manufacturing processes \cite{10.1007/978-3-540-76848-7_10}. Most traditional BPMLs follow an activity-centric paradigm.

\subsection{Need for the systematic literature review}
\label{sec:needforSLR}

Differences among notations, approaches, and methodological strategies are captured by the BPML 
 \textit{meta-models}. In fact, meta-models are used to capture the types of entities included in a notation and the way these entities can be related to each other. They can also make explicit the level of granularity of a business process (e.g., instance level, model level), or the specific sub-domain (dimension) they focus on (e.g., organisation-oriented, information-oriented, and behaviour-oriented). 
Moreover, meta-models are also defined in literature independently of
 specific BPMLs with the aim of ``navigating'' across the different BPMLs, bridge the gap across them, foster a common ground across different notations, and promote interoperability, thus further increasing their overall number. 
Meta-models are therefore a crucial source of knowledge on the constructs and rules that state-of-the-art literature considers relevant to describe business processes, yet a framework that categorise and provides a general rationale of all the meta-models described in literature is still absent. 


Indeed, while several SRLs and surveys on Model Driven Engineering (MDE), and Model Driven Architecture (MDA) exist (see e.g., \cite{Silva2015ModeldrivenEA, Santiago2012ModelDrivenEA, Loniewski2010ASR, Gonzlez2014FormalVO, Nguyen:2015:ESR:2831506.2831600}), a SRL on the different types of meta-models available in the field of BPM is still lacking. 
This lack of descriptive categorisation has several negative consequences: the first, and obvious one, is the lack of a comprehensive and easily accessible overview of what has been produced so-far in literature; the second consequence is the danger of over production of \textit{quasi same} meta-models across the community of Business Process Management (BPM); a third consequence is the lack of a framework 
 to categorise and compare the different proposals, which can act as a comprehensive common ground where to place new  proposals of meta-models; and, finally, an investigation is missing on the characteristics, strengths and limits of the current meta-models, so as to identify gaps that may originate further investigations. 

Thus, we identify the needs for this SLR in (i) the absence of a systematic study on the meta-models developed in the field of BPM, as well as, (ii) the need of the identification of dimensions where to categorise and compare meta-models and the elements they contain.  

\subsection{The research questions}
\label{sec:RQ}

Starting from the needs identified and described in the previous section we have formulated four research questions that motivate and guide our investigation. They are:
\quad \begin{description}
	\item[RQ1.] What types of business process meta-models are being proposed in literature and how can we characterise and categorise them?
	\item[RQ2.] What are the business process elements recurring across business process meta-models?
	\item[RQ3.] What is the role of a business process meta-model? 
	\item[RQ4.] Are the proposed business process meta-models evaluated? How? 
\end{description}

\rqone focuses on the differences among BPML meta-models and aims at investigating them. It also aims at identifying which are the relevant characteristics that meta-models share or in which they differ.

\rqtwo is devoted to the identification of the elements and components of business processes that
 occur in meta-models. Besides providing a photograph of the different components, this research question aims at investigating which are the elements of a business process that are (more) often represented in meta-models and whether these elements correspond to the ones that often occur in the definition of a business process. 


\rqthree is devoted to the identification and classification of the purpose for which the meta-models were introduced / used in the investigated works.


Finally, \rqfour aims at investigating the way the proposed meta-models are evaluated. This question lies on two different motivations. The first, obvious one is to map how meta-models of business processes are evaluated; the second is to assess the  importance provided to the evaluation of meta-models in different studies and to identify suggestions for possible evaluation methodologies. Indeed in literature there is a lack of guidelines and evaluation criteria for the development of meta-models in the area of business process models and this can hamper their perceived usefulness and (practical) adoption. 


\subsection{The protocol of review}
\label{sec:protocol}

The protocol of review was designed around four main phases: (i) data source and strategy; (ii) inclusion and exclusion criteria; (iii) development of the quality assessment; and finally (iv) data extraction strategy and analysis.


\subsubsection{Data source and strategy}
\label{sec:queries}
In this phase we did plan the paper repositories and search queries to be used in our SLR. We decided to perform two different types of searches. First, we decided to target paper repositories, and retrieve papers by means of keyword-based queries. Second, we decided to target proceedings of relevant conferences.

The paper repositories we decided to target are DBLP\footnote{\url{https://dblp.uni-trier.de/}}, Scopus\footnote{\url{https://www.scopus.com/search/form.uri?display=basic}}, and Web of Science\footnote{\url{https://login.webofknowledge.com/error/Error?PathInfo=2F&Error=IPError}} (WoS). 
Scopus and WoS were considered because of their extensive coverage on well established scientific literature, especially journal papers. DBLP was included because of its extensive coverage of papers in computer science including papers published in peer reviewed conference and workshop proceedings.     
To formulate the keyword-based query we queried the three paper repositories in an iterative manner that considered several combinations of keywords (e.g., process, process model, business process, business process modelling languages, meta-model, metamodel) connected by the logical operators AND and OR. The result was the adoption of the query 
\small{\begin{equation}
	\label{eq:query}
	\text{\texttt{metamodel OR meta-model AND business process OR process model}}
\end{equation}}
whose actual implementation in the syntax of the three repositories is shown in Table~\ref{table:query}. 
 
\begin{table}[t]
\centering
\scalebox{0.9}{
\begin{tabular}{lm{11cm}}
\toprule
 Scopus& \begin{minipage}{9.5cm}
  		\texttt{(``metamodel'' OR ``meta-model'') AND (``business process'' OR ``process model'')}
	  \end{minipage} \\ 
 \addlinespace[7pt]
 DBLP& \begin{minipage}{6.5cm}
 		\texttt{metamodel\textbar meta-model AND business process\textbar process model}  
	\end{minipage}\\
 \addlinespace[7pt]
 WoS& \begin{minipage}{11.5cm}
 		\texttt{((TS =``metamodel'' OR TS=``meta-model'') AND \\
		(TS=``business process'' OR TS=``process model'')) AND LANGUAGE:(English)}
	\end{minipage}\\ \bottomrule
\end{tabular}
}
\caption{Key-words on Scopus, DBLP, and WoS.}
\label{table:query}
\end{table}

The proceedings we did include in the data sources are the ones of the two reference conference venues in the BPM research area, namely the \textit{Business Process Management} (BPM) conference series\footnote{\url{https://link.springer.com/conference/bpm}} and the  \textit{Conference on Advanced Information Systems Engineering} (CAiSE) series\footnote{\url{https://link.springer.com/conference/caise}}. 


\subsubsection{Inclusion and Exclusion criteria} 
\label{sub:inclusion_and_exclusion_criteria}

The next step of the protocol was to define some relevant criteria in order to evaluate the appropriateness of the papers returned as query results for this study and thus filter them.  

\begin{table}
\centering
\scalebox{0.9}{
\begin{tabular}{lp{11cm}}
\toprule
IC 1:& The paper proposes a meta-model of business processes or BPMLs.\\
IC 2:& The meta-model is either originally developed or originally adapted by the authors.\\
IC 3:& The paper focuses mainly / exclusively on business process aspects. \\  \midrule
EC 1:& The paper is not available.\\
EC 2:& The paper is duplicate.\\
EC 3:& The paper does not belong to the BPM field.\\
EC 4:& The paper does not mainly consider the business process view, but rather it is \\
     &focused on organisational$\backslash$entrepreneurial aspects without touching the   business process level.\\
EC 5:& The paper either was not under peer-review, or it is a technical report.\\
EC 6:& The paper is almost the ``same copy" of others of the same author(s).\\
EC 7:& The paper either does not include a wide analysis of related works or does not positioned\\
     & in the state of the art.\\
EC 8:& The paper is not long enough to present a complete meta-model.\\
\bottomrule
\end{tabular}}
\caption{Inclusion and Exclusion Criteria.}
\label{table:iec}
\end{table}

Inclusion (IC) and Exclusion (EC) criteria are reported in Table~\ref{table:iec}. In order to be included papers had to satisfy all inclusion criteria IC 1 -- IC 3. Moreover, they were excluded if they did satisfy at least one of the exclusion criteria between EC 1 and EC 8. Basically, all these inclusion and exclusion criteria focus on removing duplicate, incomplete or not scientifically valid papers or refer to the primary criterion of this review, i.e.,
 the paper has to present a meta-model of business processes. Moreover, to maintain the SLR focused, and the amount of papers manageable, we restricted ourselves only to papers where the business process aspect is the main / exclusive focus of the paper, thus excluding papers mainly devoted to enterprise (meta-)models or service oriented (meta-)models.  
In this phase we decided not to consider ECs limiting the papers selection according to the date of publication. The reason for this choice lies in the fact that this is the first SLR in this field. Thus, we felt we had to consider the maximum number of papers available in literature. 





\subsubsection{Quality assessment}
\label{sec:qa}

The four quality assessment criteria we planned and used in this SLR are: 
\begin{itemize}
	\itemsep=-\parsep
 \item QA1: Is a well-defined methodology used?
\item QA2: Is the study clearly positioned within the state-of-the-art landscape?
 \item QA3: Is the goal of the study elucidated?
 \item QA4: Was the study evaluated$\backslash$validated?
\end{itemize}
We decided to use QA1--QA4 to mark papers with three possible scores: \textit{Yes (Y)}, \textit{No (N)}, and \textit{Partially (P)}, weighted 1, 0 and 0.5 respectively. A description of how QA1--QA4 were used to mark papers can be find in the file called ``quality\_assessment.pdf" at the following link: \url{https://drive.google.com/drive/folders/1_mdJBCtfQg2triqUb01AoMu7OBfahIZz}. 


\subsubsection{Data extraction strategy and analysis}
\label{sec:selectionStrategy}

\begin{figure}[b]
  \centering
    \includegraphics[width=.9\textwidth]{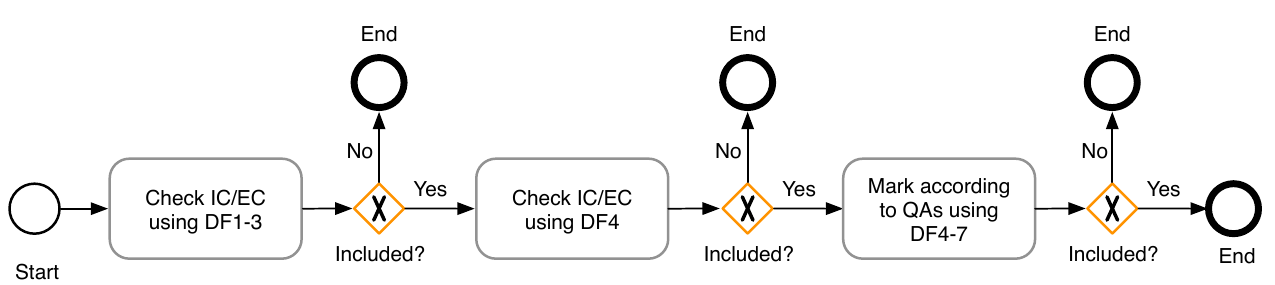}
  \caption{The selection of the primary studies}
  \label{fig:IMG_SelectionStrategy}
\end{figure}

Within this step we did plan both the data fields of the papers that were used in order to select the \emph{primary studies}, i.e., the studies analysed for addressing the research questions of the review, and the exact procedure for selecting them. 
 The outcome of this phase was a list of data fields to be used for the selection process, contained in Table~\ref{table:DFs}, and the procedure to select the primary studies, graphically illustrated in Figure~\ref{fig:IMG_SelectionStrategy}. The procedure is composed of three steps: (1) all candidate papers must be evaluated against the IC/EC exploiting just title, authors, abstract, and keywords (when present); (2) the the IC/EC are evaluated more carefully on remaining papers using the entire content of the paper; (3) the candidate primary studies are marked using the four quality assessment criteria described in Section~\ref{sec:qa} and are included in the primary studies whenever they score at least 2.5 out of the maximum possible score of 4. 

\begin{table}
\centering
\scalebox{0.9}{
\begin{tabular}{l@{\qquad\qquad}l}
\toprule
DF1: title and authors & DF2: abstract\\
DF3: keywords & DF4: content\\
DF5: related works & DF7: citations\\
\multicolumn{2}{l}{DF6: publication type (journal, conference, workshop, and book)}\\
\bottomrule
\end{tabular}}
\caption{Data Fields.}
\label{table:DFs}
\end{table}

\section{Extraction of the primary studies}
\label{sec:results}

This section briefly describes the extraction of the primary studies according to the plan presented in the previous section, and the outcomes of each single step in the process (data search, application of inclusion and exclusion criteria, and quality assessment). While the planning was conducted in March 2018, the extraction of the primary studies and of the data necessary for answering the research questions was performed \grtext{in two phases: the extraction from the paper repositories occurred
 between April 2018 and June 2018, and the extraction from the BPM and CAiSE proceedings took place between November 2018 and January 2019}.

\paragraph{Conducting the data search}

\begin{table}
\centering
\scalebox{.8}{
\begin{tabular}{lS[table-number-alignment = right]S[table-number-alignment = right]ccS[table-number-alignment = right]}
 	  & {Query} &  \multicolumn{1}{c}{No} &  &  &  \multicolumn{1}{c}{In Primary}\\
  {Source} & {Results} &  \multicolumn{1}{c}{Collections}  & & &   \multicolumn{1}{c}{Studies}  \\
 \cmidrule{1-3} \cmidrule{6-6} 
 Scopus & 1005  & 913 &    & &  31 \\
 WoS    & 367   & 367 &  &  &  16 \\
 DBLP 	& 26 	& 26   & &   & 5 \\
 \cmidrule{1-3} \cmidrule{6-6} 
 CAiSE & & 1065 & No & After & 4\\
 BPM & & 452 & Duplicates & IC/EC & 0\\
 \midrule
 Total  &  & \grtext{2823}  & \grtext{2463} &   \grtext{36} &   \\ \\
 
\end{tabular}
}
\caption{Query results and selection of Primary Studies.}
\label{table:searchresults}
\end{table}

Papers were selected using the keyword-based queries in early April 2018. Their numbers are reported in the first column of Table~\ref{table:searchresults}. 1398 papers were returned (1005 from Scopus, 367 from WoS, and 26 from DBLP), which were reduced to 1306 after the deletion of collections (e.g., entire proceedings) which were not considered as a single item in this survey. All 452 papers from the BPM conferences (starting from 2003 to 2018) and all 1065 papers published in the CAiSE conferences (starting from 1990 to 2018) were also included in the initial set of papers to be considered\footnote{We have not considered papers related to keynotes speeches and tutorials from both the BPM and CAiSE proceedings.}.
The resulting \grtext{2823} papers were pruned from duplicates (papers appearing more than once in the same data source or in at least two data sources) and retracted articles thus reducing the total number of candidates to \grtext{2463}.
\footnote{Details of all the retrieved papers, and of the ones removed in each step can be found in the CSV (Comma Separated Values) files
 accessible starting from the folder at \url{https://drive.google.com/drive/folders/1_mdJBCtfQg2triqUb01AoMu7OBfahIZz?usp=sharing}}

\paragraph{Applying the inclusion and exclusion criteria}
 
\begin{figure}[h]
  \centerline{	
    \includegraphics[width=.4\textwidth]{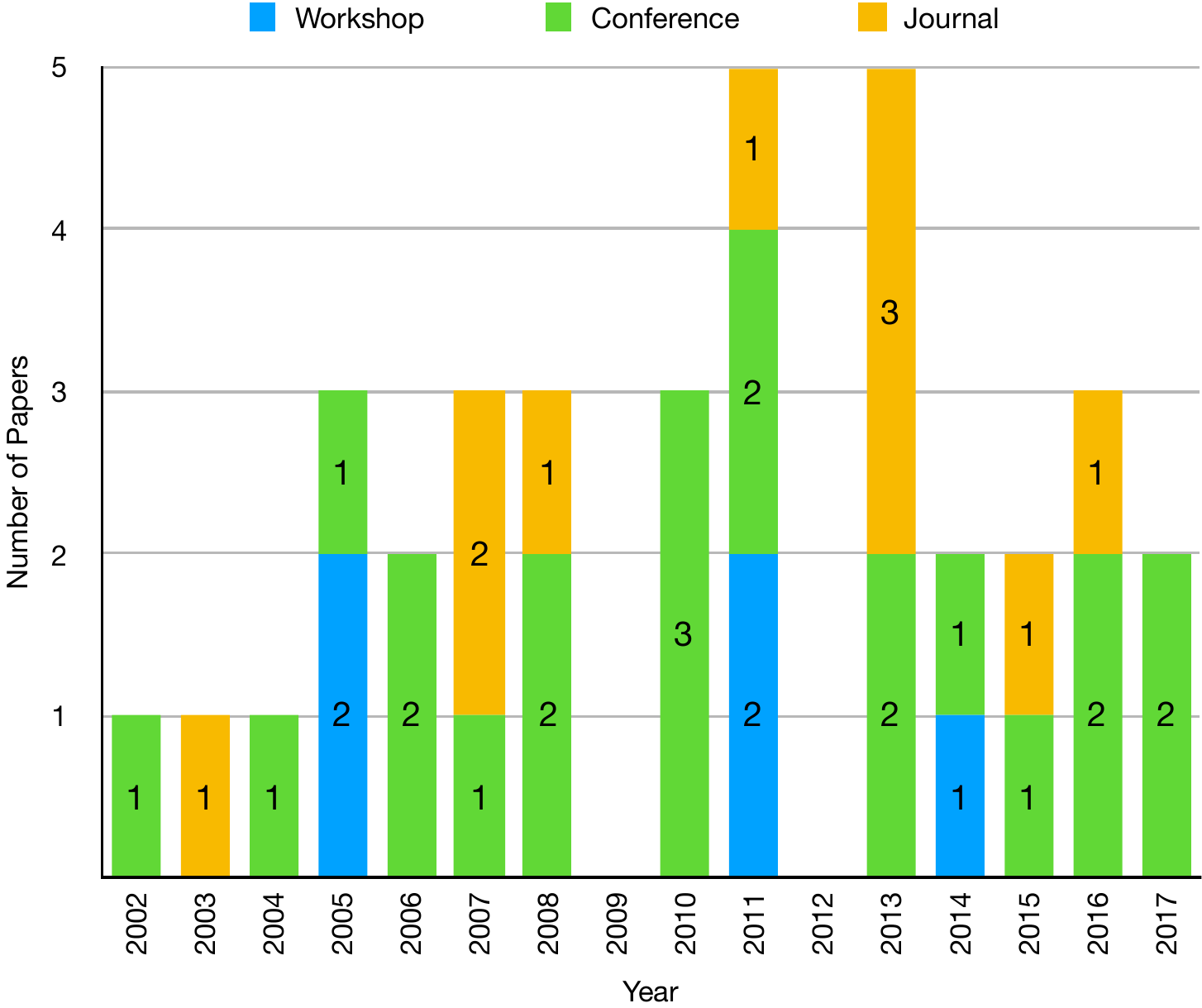}
	}
  \caption{A distribution of the primary studies along the years.}
  \label{fig:Pictures_PSdistributionyear2}
\end{figure}

The next step was to apply the IC/EC described in Table~\ref{table:iec} to the \grtext{2463} 
 papers that constitute our starting data collection. 
 As a result of this step, \grtext{36} papers were retained. These \rtext{36} papers constitute our primary studies and are listed in Table~\ref{table:primarystudieslist} classified as workshop, conference (symposium), and journal publications. Their distribution per year is reported in Figure~\ref{fig:Pictures_PSdistributionyear2}, while their venue of publication is reported in~\ref{app:sourcelist}. 

\newcommand{\gap}{.2cm}
 \begin{table}[tbp]
 \centerline{
 \scalebox{.8}{
 \begin{tabular}{llll}
 \toprule
  Year & Workshop Reference & Conference Reference & Journal Reference\\ 
  \midrule
  2002 &					& \citet{DBLP:conf/caise/SoderstromAJPW02} \\[\gap]
  2003 &  					& 						& \citet{DBLP:journals/jkm/PapavassiliouM03}\\[\gap]
  2004 & 					& \citet{DBLP:conf/adbis/MomotkoS04} \\[\gap]
  2005 &  \citet{DBLP:conf/bpm/GrangelCSP05} & \citet{DBLP:conf/caise/RussellAHE05} \\ 
  	   & \citet{thom2005improving}\\[\gap]
  2006 &  					& \citet{DBLP:conf/sac/ListK06}\\
  	   & 				    & \citet{DBLP:conf/caise/WeigandJABEI06} \\[\gap]
  2007 &  					& \citet{DBLP:conf/iceis/KorherrL07}	& \citet{DBLP:journals/ijbpim/AxenathKR07}\\
  	   & 					& 	& \citet{Farrell2006FormalisingW}\\[\gap]
  2008 & 				    & \citet{DBLP:conf/ecmdafa/HolmesTZD08}& \citet{DBLP:journals/ijbpim/RosemannRF08}\\
       &                    & \citet{DBLP:conf/er/RosaDHMG08} \\[\gap]
  2010 & 					& \citet{DBLP:conf/dexa/NicolaMPS10} \\ 
       & 					& \citet{DBLP:conf/icis/HuaZS10}\\
       & 					& \citet{DBLP:conf/sac/SantosAG10}\\[\gap]
  2011 & \citet{DBLP:conf/caise/HeidariLK10} 	& \citet{DBLP:conf/edoc/BruningG11} & \citet{DBLP:journals/infsof/StrembeckM11}\\
       & \citet{DBLP:conf/bpmn/Natschlager11} & \citet{DBLP:conf/hicss/WeissW11a} \\[\gap]
  2013 &					& \citet{DBLP:conf/iceis/BouneffaA13} & \citet{DBLP:journals/jodsn/CherfiAC13}\\
  	   & 					& \citet{DBLP:conf/wecwis/HeidariLBB13}	& \citet{DBLP:journals/is/DamaggioHV13}\\
       & 					& 														& \citet{DBLP:journals/scp/MosserB13}\\[\gap]
  2014 & \citet{DBLP:conf/wise/KunchalaYY14} & \citet{DBLP:conf/caise/RuizCEFP14}	\\[\gap] 
  2015 & 					& \citet{DBLP:conf/es/SprovieriV15}	  & \rtext{\citet{martins2015business}}\\[\gap]
  2016 & 					& \citet{DBLP:conf/isda/HassenTG16} & \citet{DBLP:journals/infsof/ArevaloCRD16}\\
	   & 					& \citet{krumeich2016modeling}\\[\gap]
  2017 & 					& \citet{hassen2017extending} & \\
	   & 					& \citet{DBLP:conf/wi/DorndorferS17} \\
\bottomrule	   
\end{tabular}
}}
    \caption{The Primary Studies.}
    \label{table:primarystudieslist}
\end{table}

As summarised in the last column of Table~\ref{table:searchresults}, 31 of these 36 papers were extracted (at least) from Scopus, 16 (at least) from WoS, 5 (at least) from DBLP, 4 (at least) from CAiSE. 

\paragraph{Performing the quality assessment}
\label{QAr}

\begin{table}[ht]
\begin{minipage}[b]{0.5\linewidth}
\centering
\includegraphics[width=60mm]{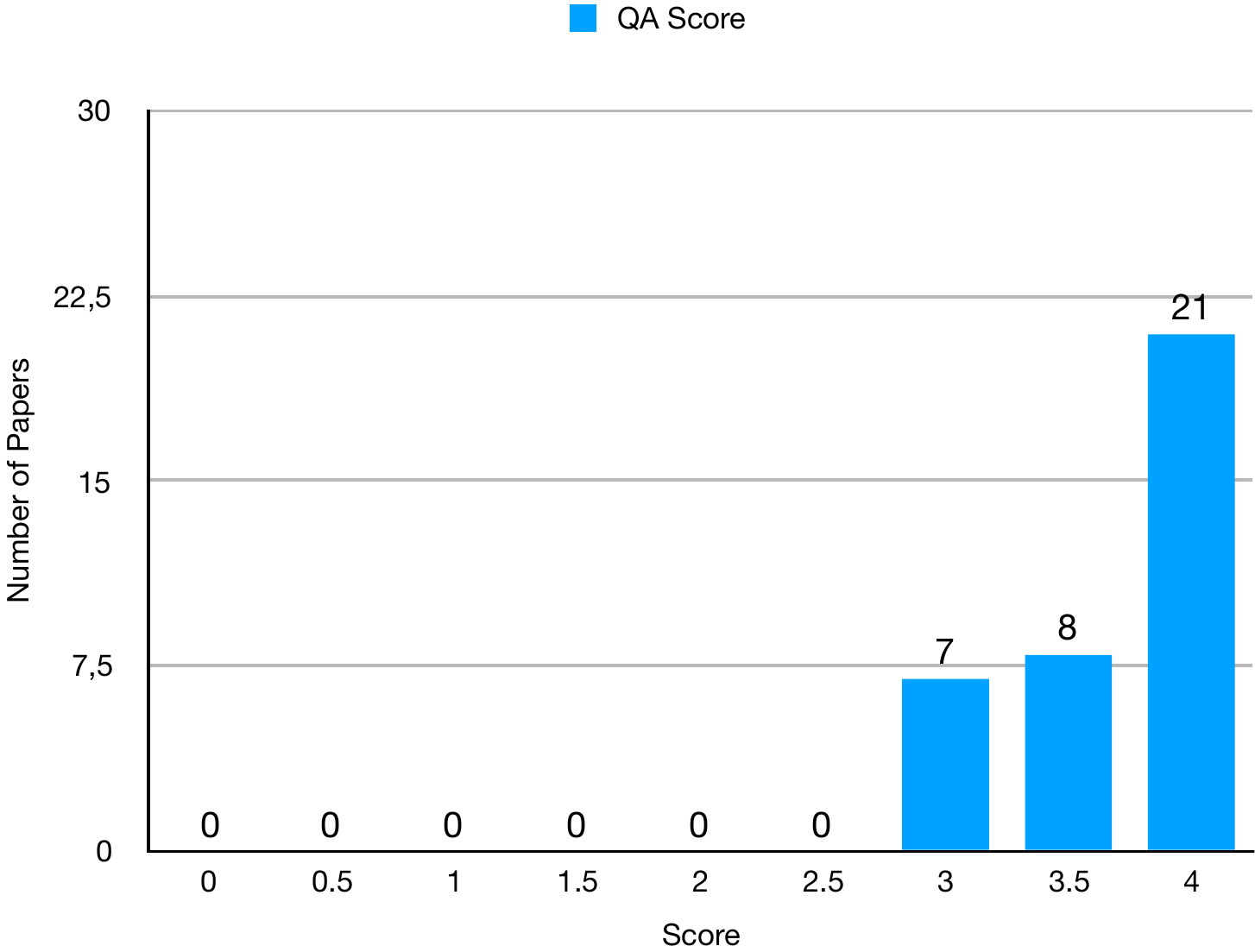}
\captionof{figure}{Results grouped by paper.}
\label{QAresults:papers}
\end{minipage}\hfill
\begin{minipage}[b]{0.56\linewidth}
\centering
\scalebox{.7}{
\begin{tabular}{rrS[table-number-alignment = right]rr}
\toprule
    & \multicolumn{1}{c}{\textbf{QA1}} & \multicolumn{1}{c}{\textbf{QA2}} & \multicolumn{1}{c}{\textbf{QA3}} & \multicolumn{1}{c}{\textbf{QA4}} \\ 
\cmidrule{2-5}
Yes & \grtext{36} & \grtext{34} & \grtext{36} & \grtext{23} \\
Partially & 0 & 2 & 0 & 6\\ 
No & 0 & 0 & 0 & 7\\
\bottomrule 
 Total Score & \grtext{36} & \grtext{36} & \grtext{36} & \grtext{36}
\end{tabular}}
    \caption{Results grouped by QA.}
    \label{QAresults:qa}
\end{minipage}
\end{table}

A summary of the quality assessment evaluation is reported in Figure~\ref{QAresults:papers} and Table~\ref{QAresults:qa}. All papers scored high on most of the questions, with \grtext{21} papers scoring \emph{Yes} in all four questions, 8 papers scoring 3.5 in total and \grtext{7} papers scoring 3 in total (see Figure~\ref{QAresults:papers}). The only \emph{No} answers did concern the evaluation, where \grtext{7} papers out of \grtext{36} had a negative score as they did not report any evaluation (see Table \ref{QAresults:qa}).   

\section{A Brief Summary of the Primary Studies}
\label{sec:summaryofpapers}
In this section we report a concise description of the papers included in the primary studies. For the sake of readability the summary is structured in three parts. First, we describe all papers presenting a meta-model of a generic business process model with no reference to a specific business process modelling language; second, we describe papers which provide meta-models specific to a given business process modelling language; and third, we describe papers that propose meta-models which were originally developed for modelling paradigms/languages that are not specific for business process modelling and were subsequently adapted to the BPM scenario.

\subsection{Primary Studies proposing language-independent meta-models} 
\label{sec:general_meta_models}


A first group of papers (\cite{DBLP:conf/wecwis/HeidariLBB13,DBLP:conf/sac/ListK06,DBLP:conf/caise/SoderstromAJPW02,DBLP:conf/wise/KunchalaYY14}) propose general meta-models that are used to compare, integrate, translate, or evaluate specific business process modelling languages. In particular, \citet{DBLP:conf/wecwis/HeidariLBB13} propose a meta-model abstraction obtained by integrating concepts from different business process modelling languages that is then used to classify and
 compare
 different process modelling languages.
Similar efforts are contained in \cite{DBLP:conf/sac/ListK06, DBLP:conf/caise/SoderstromAJPW02} where rich meta-models that encompass the typical behavioural perspective are provided. While these four papers consider business process modelling languages that follow the so-called procedural approach to business process modelling, the work of \citet{DBLP:conf/wise/KunchalaYY14}  exploits the meta-model contained in the BALSA framework~\cite{Bhattacharya09adata-centric} to provide a comparative review of modelling languages, with special emphasis to the ones that follow the so-called artifact-centric modelling approach to business process.

Another group of papers (\cite{DBLP:conf/caise/HeidariLK10,DBLP:conf/icis/HuaZS10,DBLP:journals/jodsn/CherfiAC13,thom2005improving}) exploits meta-models to foster the quality of business process models. 
\citet{DBLP:conf/caise/HeidariLK10} introduce a general meta-model of a business process, inspired by a set of specific business process modelling languages, and they enrich
 with different quality related information connected to the relevant modelling constructs. Instead, \citet{DBLP:conf/icis/HuaZS10} and \citet{DBLP:journals/jodsn/CherfiAC13} address  the issue of improving the quality of a business process model by exploiting domain knowledge. Both works aim at integrating the meta-models of domain ontologies to the one of a business process. Moreover, \citet{DBLP:conf/icis/HuaZS10} present a domain ontology based procedure towards business process modelling, while \citet{DBLP:journals/jodsn/CherfiAC13} describe a set of general mappings between the two meta-models, and their instantiation in the Object Constraint Language (OCL). Finally, \citet{thom2005improving} aim at fostering the quality of business processes via the usage of business (sub)process patterns, and introduce a meta-model called ``Transactional Metamodel of business process'', whose aim is to guide the definition and usage of patterns based on task flow descriptions as well as organisational structural aspects. The meta-model also supports the generation of patterns through BPEL4WS, Business Process Execution Language for Web Services. 

Moving from quality to flexibility, the work of \citet{DBLP:journals/ijbpim/RosemannRF08} proposes the use of a meta-model to represent relevant contextual information in business processes as a way to improve their flexibility and  adaptability. \citet{DBLP:conf/iceis/BouneffaA13} focus on the management of change in software applications based on business process models.
  The work proposes a meta-model of business processes extended with a taxonomy of business process change operations. The paper also provides an implementation using a software change management platform based on a set of the Eclipse Workbench plug-ins.

A further set of papers (\cite{DBLP:conf/isda/HassenTG16,DBLP:journals/ijbpim/AxenathKR07,martins2015business,DBLP:conf/edoc/BruningG11}) provides rich meta-models which cover different dimensions related to business processes. In particular, the work of \citet{DBLP:conf/isda/HassenTG16} focuses on knowledge-intensive business processes and proposes a rich meta-model for business processes which covers the functional, organisational, behavioural, informational, intentional and knowledge dimensions. The meta-model is then applied to a real world medical scenario. Similarly, \citet{DBLP:journals/ijbpim/AxenathKR07} propose AMFIBIA, a business process meta-model covering a wide set of static and dynamic aspects of processes together with their interactions. The meta-model is also used to realise a formalism-independent workflow engine with the same name. The work in \citet{martins2015business} presents a meta-model that enriches the typical constructs of business process modelling languages with layers and concepts coming from work practice information, and in particular with service, structure, and activity related concepts. Finally, the work in \citet{DBLP:conf/edoc/BruningG11} presents a meta-model expressed in the UML and OCL languages oriented to the representation of workflows in a declarative manner. The meta-model covers the behavioural, data-related and organisational aspects of workflows. 

\citet{DBLP:conf/caise/WeigandJABEI06} extend the behavioural view of business processes with the notion of value. In particular their work
 provides an analysis on the notion of value (objects) in the context of business processes, or, more specifically, in the context of the activities involved when transferring value objects between business actors. 
  
\citet{DBLP:journals/jkm/PapavassiliouM03} aim at integrating knowledge and process management, focusing on knowledge-intensive processes. The work
 presents an approach for the integration of knowledge tasks and knowledge objects in business process models fostered by meta-models. 
 
Finally, the work in \citet{DBLP:conf/caise/RussellAHE05} encompasses the typical behavioural view on business process modelling languages by adding a resource perspective. In order to present modelling patterns that involve the control flow dimension together with the resource one, the authors first present a rich description of workflow and resource concepts, including the relations that hold among them, which constitute a de-facto meta-model.


\subsection{Primary Studies proposing language specific meta-models} 
\label{language_specifMM}


This group of papers can be further divided into papers focusing on well-known business process modelling languages (\cite{DBLP:journals/infsof/ArevaloCRD16, hassen2017extending,DBLP:conf/wi/DorndorferS17,DBLP:conf/bpmn/Natschlager11,DBLP:conf/iceis/KorherrL07,krumeich2016modeling,DBLP:conf/sac/SantosAG10,DBLP:conf/er/RosaDHMG08,DBLP:conf/es/SprovieriV15,DBLP:journals/is/DamaggioHV13,Farrell2006FormalisingW,DBLP:journals/infsof/StrembeckM11,DBLP:conf/caise/RuizCEFP14}) or on a novel language proposed in the paper, or in related papers by the same authors, together with its meta-model (\cite{DBLP:conf/adbis/MomotkoS04,DBLP:conf/dexa/NicolaMPS10,DBLP:conf/hicss/WeissW11a}).  

Within the group tailoring popular business process modelling languages, a conspicuous number of papers (\cite{DBLP:journals/infsof/ArevaloCRD16, hassen2017extending,DBLP:conf/wi/DorndorferS17,DBLP:conf/bpmn/Natschlager11}) refer to meta-models for BPMN or for BPMN extensions. 
The work of \citet{DBLP:journals/infsof/ArevaloCRD16} proposes to extend BPMN 2.0 with a time related perspective. In particular it provides a  taxonomy of declarative rules based on (1) a BPMN meta-model extension that incorporates the time dimension, and (2) an OCL (Object Constraint Language) formalisation of the time related declarative rules. An example of application of how the proposed time-based extension can help in the extraction of business processes from legacy databases is also provided.  
\citet{hassen2017extending} present BPMN4KM, a BPMN 2.0 extension that
 focuses on the knowledge dimension for Sensitive Business Processes (SBPs). The extension is done by exploiting BPM4KI, a business process independent generic meta-model common to current BPM formalisms. 
%
%
\citet{DBLP:conf/wi/DorndorferS17} provide a BPMN 2.0 meta-model extension tailored to the business processes executed in mobile contexts, that is, business processes supported by mobile devices/applications. 
The last paper of this group provides a BPMN meta-model in the form of formal ontologies. \citet{DBLP:conf/bpmn/Natschlager11} provides a formal ontology for BPMN 2.0, together with some examples of usage as a knowledge base and as syntax checker for BPMN 2.0 models. 


The second most popular language in our primary study is EPC (Event-Driven Process Chain)~\cite{EPC-Davis}, which is investigated in \cite{DBLP:conf/iceis/KorherrL07,krumeich2016modeling,DBLP:conf/sac/SantosAG10,DBLP:conf/er/RosaDHMG08}. Actually, \citet{DBLP:conf/iceis/KorherrL07} address both BPMN and EPC. In fact, the paper
 provides an extension of both the EPC and the BPMN meta-models that
 adds the concepts of process goals and performance measures. 
\citet{krumeich2016modeling} aim at modelling complex event patterns in EPC and automatically transform them into an executable Event Pattern Language (EPL). The modelling of the complex event patterns exploits an extension of the EPC meta-model proposed in the paper, together with a modelling technique incorporated in the ARIS Business Process Analysis Platform. 
ARIS EPCs are also the focus of the work of~\citet{DBLP:conf/sac/SantosAG10}. This paper presents an 
ontological analysis of the EPC business process modelling notation supported in the ARIS Toolset. The ontological analysis makes use of the Unified Foundational Ontology (UFO)~\cite{Guizzardi-Wagner-UFO} and of a further meta-model of the ARIS Method, introduced by the same authors in \cite{IJBPIM-ARIS}. It provides a precise real-world semantics for business process models represented through EPCs as supported by the ARIS Toolset.
The last paper in this group is proposed by~\citet{DBLP:conf/er/RosaDHMG08}, who provide an extensive meta-model for configurable processes with advanced features for capturing resources involved in the performance of tasks  as well as flow of data and physical artefacts. While being potentially applicable to other notations, the meta-model is defined as an extension of EPCs. 

Two further papers (\cite{DBLP:conf/es/SprovieriV15,DBLP:journals/is/DamaggioHV13}) refer to the so-called artefact-centric approach to business process modelling. 
\citet{DBLP:conf/es/SprovieriV15} propose an algorithm to support the run-time planning of partly structured parts of a business process modelled in the CMMN (Case Management Model Notation) modelling language. The run-time planning is used to find an appropriate sequence of tasks. The selection and specification of tasks is supported by an extension of the CMMN meta-model. The work of~\citet{DBLP:journals/is/DamaggioHV13} is instead focused on the Guard-Stage-Milestone (GSM) formalism. The GSM meta-model is introduced together with three different, and provably equivalent, formulations of the GSM operational semantics.

The last papers addressing existing specific BPM languages are \cite{Farrell2006FormalisingW}, \cite{DBLP:journals/infsof/StrembeckM11} and \cite{DBLP:conf/caise/RuizCEFP14}, which focus on YAWL, UML2 activity models, and Communication Analysis, respectively.
\citet{Farrell2006FormalisingW} provide a formal specification of business process workflows. The authors start by representing business processes in terms of a meta-model called \emph{Liesbet}, which
 is based on YAWL patterns, and then formally characterise Liesbet using Milner's Calculus of Communicating Systems (CCS).
The work of \citet{DBLP:journals/infsof/StrembeckM11} aims at combining business processes and role-based access control (RBAC) models. To reach this goal they propose a general and language-independent formal meta-model for process-related RBAC models and they also instantiate this meta-model as an extension for UML2 activity models.  
\citet{DBLP:conf/caise/RuizCEFP14} aim to align and integrate a goal-oriented modelling language, namely i*,  and a business process-oriented modelling language, namely Communication Analysis, through a reference ontology called FRISCO. The authors also illustrate guidelines for a top-down usage of the two languages, and a tool to support the approach.

We conclude with 3 papers presenting meta-models that are used to introduce novel business process modelling languages (\cite{DBLP:conf/adbis/MomotkoS04,DBLP:conf/dexa/NicolaMPS10,DBLP:conf/hicss/WeissW11a}). 
\citet{DBLP:conf/adbis/MomotkoS04} present the business process query language BPQL. To do this, a meta-model of workflows is introduced to represent the workflow model upon which BPQL has to operate. 
\citet{DBLP:conf/dexa/NicolaMPS10} present a platform for business process modelling and verification. The platform is centred around the logic-based language BPAL (Business Process Abstract Language), introduced by some of the
 authors in \cite{DBLP:conf/iicai/NicolaLM07}. BPAL mainly focuses on the control flow perspective of business processes and is illustrated in the paper together with its meta-model.  
Finally, \citet{DBLP:conf/hicss/WeissW11a} introduce the semantic process modelling language SBPML, together with its meta-model. SBPML is a domain-specific language tailored to the financial sector and thus, its meta-model contains domain-independent elements as well as domain-specific ones describing financial processes related elements in all the process, organisation, data object and resource views. 

\subsection{Primary Studies on non business process modelling languages} 
\label{other_paradigmsMM}


In this group we present papers that refer to modelling languages or approaches that do not strictly pertain to the business process modelling one (\cite{DBLP:conf/ecmdafa/HolmesTZD08,DBLP:journals/scp/MosserB13,DBLP:conf/bpm/GrangelCSP05}). In particular we present here meta-models that refer to languages developed in neighbouring areas such as the ones of Service Oriented Computing (SOA) or Enterprise modelling. While referring to languages originally developed for something different from business processes, the papers were selected as the business process aspect is the main / exclusive focus of the paper.  

The first two papers in this group refer to SOA languages. The work proposed by \citet{DBLP:conf/ecmdafa/HolmesTZD08} provides a meta-model that supports the modelling of business processes that involve humans. Then this general abstract human meta-model is refined towards a technology-specific one for which a model-to-code transformation will be defined in order to obtain a BPEL4People process.  
The work of \citet{DBLP:journals/scp/MosserB13} instead, presents a new logic-based modelling language called ADORE, whose meta-model is inspired by the SOA language BPEL, and whose aim is to enable process designers modelling and composing (fragments of) business processes.

The work developed by \citet{DBLP:conf/bpm/GrangelCSP05} stems from the enterprise modelling area. It focuses on the enterprise modelling language meta-model POP*\cite{CHEN2009175}, and develops the specific part dedicated to the business process view of the POP* meta-model. The meta-model focuses on activities and elements needed to enact and execute processes in a collaborative enterprise. 



\section{Answering the research questions}
\label{sec:RRQs}

This section presents the answers of all the research questions introduced in Section~\ref{sec:RQ}. 

\subsection{Answering \rqone}
\label{subsec:RQ1}

In answering \rqone we aim at investigating the characteristics of the meta-models introduced in the literature and whether there is a way to categorise them. 
Obviously this question could have several answers, depending on the perspective exploited to look at the meta-models. In this paper we answer \rqone in two different steps. 

The first characterisation we did observe in looking at the papers is based on their relationship with specific modeling languages or paradigms.\footnote{This characterisation is roughly the one that we have exploited in reporting the concise description of the primary studies provided in Section~\ref{sec:summaryofpapers}. Even if the categorisation of the primary studies in different groups was obtained when answering \rqone, and will therefore be discussed here, we decided to exploit it also in Section~\ref{sec:summaryofpapers} for the sake of presentation.}  
Indeed, by looking at the meta-models of the \psnumber primary studies, we can observe that they can be divided in two mutually exclusive categories: the first one, hereafter
called \bpml, contains meta-models whose primary aim is to describe business processes; the second one, called \nobpml, 
 contains instead meta-models that
 describe business processes but whose primary aim is to describe something different from a business process (e.g., a service, an enterprise model and so on). 
These two categories, in turn, contain two different sub-categories: the first one, called \gen, which contains general meta-models of business processes that are not related to any concrete business process modelling language; the second, hereafter named \lsp, which contains meta-models of concrete business process modelling languages. In turn, \lsp can be divided in two (sub-)sub-categories: the first one, called \oldl, contains meta-models of an existing well-established business process modelling language, while the second one, hereafter called \newl, contains meta-models of new modelling language proposed in the very same paper, or by the same author in closely related papers.  

\begin{table}[ht]
\centering
\scalebox{0.8}{
\begin{tabular}{ll}
\toprule
  Category & Primary studies  \\ \midrule
   \bpml & \\
   \quad \gen & \cite{DBLP:conf/caise/SoderstromAJPW02,DBLP:conf/wecwis/HeidariLBB13,DBLP:conf/sac/ListK06,DBLP:conf/wise/KunchalaYY14,DBLP:conf/caise/HeidariLK10,DBLP:conf/isda/HassenTG16,martins2015business,DBLP:conf/edoc/BruningG11,DBLP:conf/caise/WeigandJABEI06,DBLP:journals/ijbpim/AxenathKR07,DBLP:journals/jkm/PapavassiliouM03,DBLP:conf/caise/RussellAHE05,DBLP:conf/icis/HuaZS10,DBLP:journals/jodsn/CherfiAC13,thom2005improving,DBLP:journals/ijbpim/RosemannRF08,DBLP:conf/iceis/BouneffaA13} \\
	\quad \lsp \\
	\qquad \newl & \cite{DBLP:conf/adbis/MomotkoS04,DBLP:conf/dexa/NicolaMPS10,DBLP:conf/hicss/WeissW11a}\\
	\qquad \oldl & \cite{DBLP:journals/infsof/ArevaloCRD16, hassen2017extending,DBLP:conf/wi/DorndorferS17,DBLP:conf/bpmn/Natschlager11,DBLP:conf/iceis/KorherrL07,krumeich2016modeling,DBLP:conf/sac/SantosAG10,DBLP:conf/er/RosaDHMG08,DBLP:conf/es/SprovieriV15,DBLP:journals/is/DamaggioHV13,Farrell2006FormalisingW,DBLP:journals/infsof/StrembeckM11,DBLP:conf/caise/RuizCEFP14}\\
	\nobpml\\
	\quad \gen & \cite{DBLP:conf/bpm/GrangelCSP05} \\
	\quad \lsp \\
	\qquad \newl & \cite{DBLP:journals/scp/MosserB13}\\
	\qquad \oldl & \cite{DBLP:conf/ecmdafa/HolmesTZD08}\\ \bottomrule
\end{tabular}}
\caption{A first characterisation of meta-models.}
\label{table:GLOtable}
\end{table}

Table~\ref{table:GLOtable} provides the list of these categories (where indentation is used to indicate subclasses), together with a classification of the primary studies w.r.t. the categories just introduced. In short, \grtext{18} papers present meta-models that are independent from any specific modelling language, while \grtext{18} papers belong to the language specific class \lsp. Of the latter, the biggest group is the one describing meta-models of existing business process modelling languages (\grtext{14} papers). The remaining papers describe meta-models of newly proposed business process modelling languages (3 papers), meta-models of newly proposed languages that contain business process related aspects but that
 are not specific business process modelling languages (\grtext{1 paper}) 
 and meta-models of existing modelling languages that are not specific to the business domain (1 paper). 

By looking at the primary studies we did notice further characteristics the meta-models can have, ranging from the scope of the meta-model, to the type of language used to express it, to the tool support provided in the approach.\footnote{Note that, in answering \rqone we do not take into account the process model elements described by the meta-models (e.g., whether they enable to describe roles, goals, artefacts and so on). This is due to the fact that we have a specific research question (\rqtwo) devoted to investigate
 what is described by the meta-models.} This second set of categories we did extract from the primary studies is: 

\begin{itemize}
	\itemsep=-\parsep
\item Formal (\frm): the meta-model is described by means of a formal language;
\item Meta-models of models (\model): the meta-model considers (only) the process model dimension;
\item Meta-models of executions (\exe): the meta-model considers (only) the process execution dimension 
\item Meta-models of executions and models  (\modexe): the meta-model considers both the process execution and the process model dimensions; 
\item Procedural (\proc): the meta-model adheres to a procedural view of business processes; 
\item Declarative (\dec): the meta-model adheres to a declarative view of business processes; 
\item Activity-centric (\act): the meta-model adheres to an activity-centric view of business processes; 
\item Artefact-centric (\art): the meta-model adheres to an artefact-centric view of business processes; 
\item Domain (\dm): The meta-model is domain dependent;
\item Evaluation (\eval): The meta-model is (somehow) evaluated.
\end{itemize}

\begin{table}[ht]
\centering
\scalebox{.8}{
\begin{tabular}{ll}
\toprule
  Class & Primary studies  \\ \midrule    
  \frm & \cite{DBLP:journals/infsof/ArevaloCRD16,DBLP:conf/bpmn/Natschlager11,DBLP:conf/er/RosaDHMG08, DBLP:conf/es/SprovieriV15,DBLP:journals/is/DamaggioHV13,Farrell2006FormalisingW,DBLP:journals/infsof/StrembeckM11, DBLP:conf/dexa/NicolaMPS10,DBLP:journals/scp/MosserB13} \\
  %
  %
  \model & \cite{DBLP:conf/wecwis/HeidariLBB13,DBLP:conf/sac/ListK06,   DBLP:conf/isda/HassenTG16,  martins2015business,DBLP:conf/caise/RuizCEFP14, DBLP:conf/caise/WeigandJABEI06,DBLP:journals/jkm/PapavassiliouM03,  DBLP:conf/icis/HuaZS10, DBLP:journals/jodsn/CherfiAC13,DBLP:journals/ijbpim/RosemannRF08,DBLP:journals/infsof/ArevaloCRD16,  hassen2017extending,  DBLP:conf/wi/DorndorferS17,DBLP:conf/bpmn/Natschlager11,  DBLP:conf/iceis/KorherrL07,DBLP:conf/sac/SantosAG10,DBLP:conf/hicss/WeissW11a,DBLP:conf/bpm/GrangelCSP05}\\
  \exe & \cite{thom2005improving, DBLP:conf/es/SprovieriV15, DBLP:journals/is/DamaggioHV13,DBLP:conf/adbis/MomotkoS04,  DBLP:conf/ecmdafa/HolmesTZD08,DBLP:journals/scp/MosserB13} \\
  \modexe & \cite{DBLP:conf/caise/SoderstromAJPW02,DBLP:conf/wise/KunchalaYY14, DBLP:conf/caise/HeidariLK10, DBLP:conf/edoc/BruningG11,  DBLP:journals/ijbpim/AxenathKR07, DBLP:conf/caise/RussellAHE05,DBLP:conf/iceis/BouneffaA13, krumeich2016modeling,DBLP:conf/er/RosaDHMG08,  Farrell2006FormalisingW,  DBLP:journals/infsof/StrembeckM11,DBLP:conf/dexa/NicolaMPS10} \\
	\proc & all, except	\cite{DBLP:conf/wise/KunchalaYY14,DBLP:conf/edoc/BruningG11,DBLP:conf/dexa/NicolaMPS10, DBLP:journals/infsof/ArevaloCRD16,DBLP:journals/is/DamaggioHV13,DBLP:conf/es/SprovieriV15,DBLP:journals/scp/MosserB13,DBLP:conf/caise/WeigandJABEI06}\\
  \dec & \cite{DBLP:conf/wise/KunchalaYY14,DBLP:conf/edoc/BruningG11,DBLP:conf/dexa/NicolaMPS10, DBLP:journals/infsof/ArevaloCRD16,DBLP:journals/is/DamaggioHV13,DBLP:conf/es/SprovieriV15,DBLP:journals/scp/MosserB13}\\
  \act & all, except \cite{DBLP:conf/wise/KunchalaYY14, DBLP:journals/is/DamaggioHV13,DBLP:conf/es/SprovieriV15,DBLP:conf/caise/WeigandJABEI06}	\\
  \art & \cite{DBLP:conf/wise/KunchalaYY14, DBLP:journals/is/DamaggioHV13,DBLP:conf/es/SprovieriV15}\\
  
%
%
 %
  \dm & \cite{DBLP:conf/hicss/WeissW11a, DBLP:conf/wi/DorndorferS17}  \\
  %
%

  \eval &  \cite{DBLP:conf/isda/HassenTG16,DBLP:journals/scp/MosserB13, DBLP:conf/wecwis/HeidariLBB13,DBLP:conf/icis/HuaZS10,DBLP:journals/jodsn/CherfiAC13,DBLP:journals/ijbpim/RosemannRF08,DBLP:conf/edoc/BruningG11,DBLP:conf/caise/WeigandJABEI06,DBLP:journals/jkm/PapavassiliouM03,DBLP:journals/infsof/ArevaloCRD16,DBLP:conf/iceis/KorherrL07,krumeich2016modeling,DBLP:journals/infsof/StrembeckM11,DBLP:conf/bpm/GrangelCSP05,DBLP:conf/wecwis/HeidariLBB13,DBLP:conf/sac/SantosAG10,DBLP:conf/wi/DorndorferS17,DBLP:conf/hicss/WeissW11a,DBLP:conf/bpmn/Natschlager11} \\ 
 \bottomrule
\end{tabular}}
\caption{A second characterisation of meta-models.}
\label{table:Addtable}
\end{table}

Table~\ref{table:Addtable} provides a description of the primary studies w.r.t. the classes introduced above. \grtext{9} primary studies provide a formal representation of the meta-model they describe. \grtext{Half} of the primary studies (\grtext{18}) are focused on the model dimension only, \grtext{6} consider the execution dimension only, and \grtext{12} take into account both. Concerning the approach towards business process modelling, most primary studies adhere to the traditional procedural  and activity-centric based view on business processes (\grtext{28} and \grtext{32} papers respectively), with very few papers taking a declarative or artefact-centric view.\footnote{The work of \citet{DBLP:conf/caise/WeigandJABEI06} appears to provide an original, yet uncommon, ``value centred'' approach towards business process modelling that seems to share some characteristics of artefact-centric declarative approaches. Nonetheless, a classification under the \dec and \art categories was not possible, due to a lack of details. }   

Another aspect to be taken into account is the one related to the domain (in)dependency of the meta-model. In our study, only two papers focus on domain-specific business processes, while all the others are domain-dependent. The two domains are the financial sector~\cite{DBLP:conf/hicss/WeissW11a} and a context-sensitive mobile domain~\cite{DBLP:conf/wi/DorndorferS17}. 
Finally, \grtext{slightly more than 50\%} of the meta-models are (somehow) evaluated (\eval), even if the level of evaluation differs greatly among the different papers. This aspect will be better discussed in Section \ref{subsec:RQ4}, when answering \rqfive.\footnote{Please note that QA4 did concern with an evaluation/validation of the study which could encompass the meta-model while here we refer explicitly to the evaluation of the meta-model.}

\subsection{Answering \rqtwo}
\label{subsec:RQ2}

\newcommand{\spann}{\textwidth}
\begin{table}[t]
\scriptsize
 \centering
\scalebox{.8}{
 \begin{tabular}{ll}
 \toprule
 Macro-element & Element\\
 \midrule
 \begin{minipage}{.2\textwidth}
 \textit{activity}\\
 \textbf{(\grtext{9/\TODOGRETA{64}})}
 \end{minipage} &
 \begin{minipage}{\spann}
 \TODOGRETA{\textbf{\texttt{activity}}~\textbf{(27)}}, \TODOGRETA{\textbf{\texttt{atomic activity}}~\textbf{(9)}},
 \TODOGRETA{\textbf{\texttt{compound activity}}~\textbf{(13)}}, \TODOGRETA{\texttt{activity instance}~\textbf{(4)}},
 \texttt{manual activity}~\textbf{(\grtext{2})}, \texttt{automatic activity}~\textbf{(2)},
 \texttt{collaborative organisational activity}~\textbf{(2)},
 \texttt{critical organizational activity}~\textbf{(2)}, \texttt{cancel activity}~\textbf{(3)}
 \end{minipage}
 \\
 \cmidrule{2-2}
 \begin{minipage}{.2\textwidth}
 \textit{event}\\
 \textbf{ (10/\TODOGRETA{41})}
 \end{minipage} &
 \begin{minipage}{\spann}
\TODOGRETA{\textbf{\texttt{event-EPC}}~\textbf{(4)}, \textbf{\texttt{event-BPMN}}~\textbf{(9)}}, \texttt{event sub-process}~\textbf{(3)}, \texttt{throw event}~\textbf{(2)},
 \texttt{interrupting}~\textbf{(2)}, \texttt{start event}~\textbf{(\grtext{6})}, \texttt{intermediate
 event}~\textbf{(\grtext{3})}, \texttt{end event}~\textbf{(\grtext{8})}, \texttt{message event}~\textbf{(2)},
 \texttt{event location}~\textbf{(2)}
 \end{minipage}
 \\
 \cmidrule{2-2}
 \begin{minipage}{.2\textwidth}
 \textit{state}\\
 \textbf{(5/\TODOGRETA{27})}
 \end{minipage}
  &
 \begin{minipage}{\spann}
 \TODOGRETA{\texttt{state}~\textbf{(4)}}, \TODOGRETA{\textbf{\texttt{precondition}}~\textbf{(9)}}, \TODOGRETA{\texttt{postcondition}~\textbf{(8)}},
 \texttt{data input}~\textbf{(3)}, \texttt{data output}~\textbf{(3)}
 \end{minipage}
 \\
 \cmidrule{2-2}
 \begin{minipage}{.2\textwidth}
 \textit{sequence flow}\\
 \textbf{(18/\TODOGRETA{91})}
 \end{minipage}
 &
 \begin{minipage}{\spann}
 \texttt{conditional control flow}~\textbf{(4)},
 \texttt{sequence}~\textbf{(3)}, \texttt{multimerge}~\textbf{(2)}, \texttt{multi choice}~\textbf{(2)},
 \texttt{syncronisation point}~\textbf{(2)}, \texttt{connecting object}~\textbf{(\grtext{7})},
 \TODOGRETA{\texttt{sequence flow}~\textbf{(7)}}, \texttt{condition}~\textbf{(2)}, \texttt{merge}~\textbf{(2)}, \texttt{join}~\textbf{(2)}, \texttt{fork}~\textbf{(2)},
 \TODOGRETA{\textbf{\texttt{gateway}}~\textbf{(16)}}, \texttt{complex} gateway~\textbf{(\grtext{2})}, 
 \texttt{event-based gateway}~\textbf{(\grtext{2})}, \TODOGRETA{\textbf{\texttt{parallel gateway}}~\textbf{(12)}},
 \TODOGRETA{\textbf{\texttt{inclusive gateway}}~\textbf{(9)}}, \TODOGRETA{\textbf{\texttt{exclusive gateway}}~\textbf{(11)}}, 
 \texttt{flow operator}~\textbf{(4)}
 \end{minipage}
 \\
 \cmidrule{2-2}
 \begin{minipage}{.2\textwidth}
 \textit{time}\\
 \textbf{(3/6)}
 \end{minipage} & 
 \begin{minipage}{\spann}
 \texttt{time point}~\textbf{(2)}, \texttt{cycle time duration}~\textbf{(2)},
 \texttt{temporal dependency}~\textbf{(2)}
 \end{minipage}
 \\
 \cmidrule{2-2}
 \begin{minipage}{.2\textwidth}
 \textit{data flow} \\
 \textbf{(6/\grtext{19})}
 \end{minipage} & 
 \begin{minipage}{\spann}
 \texttt{message flow}~\textbf{(\grtext{5})}, \texttt{data flow}~\textbf{(5)}, 
 \texttt{association}~\textbf{(\grtext{3})}, \texttt{conversational link}~\textbf{(2)}, 
 \texttt{knowledge flow}~\textbf{(2)}, \texttt{assignment} to an actor~\textbf{(2)}
 \end{minipage}
 \\
 \cmidrule{2-2}
 \begin{minipage}{.2\textwidth}
 \textit{data object}\\
 \textbf{(17/\TODOGRETA{48})}
 \end{minipage} & 
 \begin{minipage}{\spann}
 \textbf{\texttt{artifact}}~\textbf{(\grtext{9})}, \texttt{physical artifact}~\textbf{(2)},  
 \TODOGRETA{\texttt{data object}~\textbf{(5)}}, \texttt{message}~\textbf{(3)}, \texttt{conversation}~\textbf{(3)}, 
 \texttt{call conversation}~\textbf{(2)}, 
 \texttt{information}~\textbf{(3)}, \texttt{physical knowledge support}~\textbf{(2)}, 
 \texttt{internal knowledge}~\textbf{(2)}, \texttt{tacit knowledge}~\textbf{(2)}, 
 \texttt{external knowledge}~\textbf{(2)}, \texttt{explicit knowledge}~\textbf{(2)}, 
 \texttt{procedural knowledge}~\textbf{(2)}, \texttt{knowledge}~\textbf{(3)}, 
 \texttt{document}~\textbf{(2)}, \texttt{artifact instance}~\textbf{(2)}, 
 \texttt{data store}~\textbf{(2)}
 \end{minipage}
 \\
 \cmidrule{2-2}
 \begin{minipage}{.2\textwidth}
 \textit{actor}\\
 \textbf{(14/\TODOGRETA{72})}
 \end{minipage} & 
 \begin{minipage}{\spann}
 \textbf{\texttt{actor}}~\textbf{(\grtext{14})}, \texttt{collective agent}~\textbf{(4)}, 
 \texttt{organisation}~\textbf{(\grtext{6})}, \texttt{organisation unit}~\textbf{(6)}, 
 \texttt{human expert}~\textbf{(2)}, \texttt{internal agent}~\textbf{(2)}, 
 \texttt{external agent}~\textbf{(2)}, \texttt{client}~\textbf{(4)}, 
 \texttt{position}~\textbf{(\grtext{4})}, \grtext{\texttt{application}~\textbf{(4)}}, \TODOGRETA{\textbf{\texttt{role}}~\textbf{(15)}}, \texttt{process owner}~\textbf{(2)},    
 \texttt{process participant}~\textbf{(4)}, \texttt{person}~\textbf{(3)}
 \end{minipage}
 \\
 \cmidrule{2-2}
 \begin{minipage}{.2\textwidth} 
 \textit{resource}
 \textbf{(8/\TODOGRETA{50})}
 \end{minipage} & 
 \begin{minipage}{\spann}
 \textbf{\texttt{resource}}~\textbf{(13)}, \texttt{material resource}~\textbf{(3)}, 
 \texttt{immaterial resource}~\textbf{(3)}, \texttt{information}~\textbf{(4)}, 
 \texttt{position}~\textbf{(\grtext{4})}, \textbf{\texttt{role}}~\textbf{(\TODOGRETA{15})}, \texttt{application}~\textbf{(4)},  \texttt{process participant}~\textbf{(4)}
 \end{minipage}
 \\
 \cmidrule{2-2}
 \begin{minipage}{.2\textwidth} 
 \textit{\grtext{value}}\\
 \textbf{(2/5)}
 \end{minipage} & 
 \texttt{measure}~\textbf{(3)}, \texttt{cost}~\textbf{(2)}
 \\
 \cmidrule{2-2} 
 \begin{minipage}{.2\textwidth}
 \textit{goal} \\
 \textbf{(2/8)}
 \end{minipage} & 
 \texttt{organisational objective}~\textbf{(2)}, \texttt{goal}~\textbf{(6)} 
 \\
 \cmidrule{2-2} 
 \begin{minipage}{.2\textwidth}
  \textit{context}\\
  \textbf{(2/4)}
  \end{minipage} & 
 \texttt{context}~\textbf{(2)}, \texttt{business area}~\textbf{(2)}\\  
 \bottomrule
 \end{tabular}
 }
\caption{Recurring elements in meta-models.}
\label{elementsRQ2}
\end{table}

The aim of \rqtwo is to present an overview of the \textit{elements} involved in the primary studies' meta-models. In answering this question we 
have identified \grtext{374} single elements which have been grouped in 12 sets of recurrent constructs across the classes of meta-models. We considered as single elements only those that are not collections of other elements. For instance ``business process'', ``process'' and ``control flow'' were not included in this analysis. These 12 sets identify macro-elements that appear in the primary studies' meta-models, and are: \emph{activity},  \emph{event}, \emph{state}, \emph{sequence flow}, \emph{time}, \emph{data flow}, \emph{data object},  \emph{actor}, \emph{resource}, \emph{value}\footnote{\grtext{Although the explicit element ``value'' only occurs in one of the meta-models of the primary studies \cite{DBLP:conf/caise/WeigandJABEI06} and hence does not explicitly appear among the elements of the \emph{value} group, all the elements included in this group refer to measurable aspects related to the value of a business process. }}, \emph{goal}, and \emph{context}.

To focus our analysis on 
 central elements of business processes and exclude variants that were specific to a single meta-model, we decided to concentrate our study only to the \grtext{91} (out of \grtext{374}) elements that are considered in at least two meta-models. Note that elements labelled as \textit{events} have been classified either as events with a BPMN-like semantics, i.e., ``something that happens during the course of a process'\footnote{See: \url{https://www.omg.org/spec/BPMN/2.0/About-BPMN/}} (\ele{event-BPMN}) or as events $\grave{a}$-la EPC, i.e., in terms of pre-postconditions (\ele{event-EPC}) \cite{Mendling2008}.
These \grtext{91} elements are listed in Table~\ref{elementsRQ2}, together with their corresponding group.\footnote{Note that, five elements belong to more than one set of macro-elements. They are, \texttt{information}, \texttt{position}, \texttt{role}, \texttt{application}, and \texttt{process participant}.} For each  element we report, in round brackets, the number of 
 primary studies' meta-models in which it occurs. In some cases, elements with the same or very similar meaning had different names in the meta-models. To simplify the analysis and the reporting we have classified all the syntactic variants under only one name. The list of syntactic variants is contained in the report at 
\url{https://drive.google.com/drive/folders/1_mdJBCtfQg2triqUb01AoMu7OBfahIZz}. 
Finally, for each macro-element we also report \grtext{in round brackets} the number of corresponding elements and the total occurrences of these elements within the meta-models.  
The file ``table\_elements.pdf" included in the following link \url{https://drive.google.com/drive/folders/1_mdJBCtfQg2triqUb01AoMu7OBfahIZz} contains, instead, the correspondence between each
 element and the primary studies in which
 it appears. 

As we can see, four sets of macro-elements stand up as distinctive both in terms of different elements and in terms of overall appearance. They are: \emph{activity}, \emph{sequence flow}, \emph{data object} and \emph{actor}.  
Perhaps not surprisingly, the most articulate and recurring elements are the ones belonging to the \emph{sequence flow} group, with 18 different elements appearing \grtext{91} times in total. An interesting group is the one of \emph{data object}, where we can notice a detailed description of different types of knowledge (\grtext{17} in total) that can appear in business process model elements, even though their appearance is not as common as the one of the other three groups. The second group in terms of overall appearance is the one of \emph{activity} (\grtext{64} in total), where we can notice a big presence of the \texttt{activity} element, which is the most recurring element in all the meta-models. Interestingly enough, one of the aspects that distinguishes  business process models from other types of processes, that is, the \emph{actor}/organisational aspect, is well represented in most of the meta-models both in terms of variety of elements (14) and overall presence (\grtext{72}). Similarly interesting is the fact that key elements of goal (or value) appearing in almost all the modern definitions of business processes (such as the one of Weske provided in Section~\ref{sec:needforSLR}) have instead a very low \grtext{(or in some cases just implicit)} presence in business process meta-models. 
When it comes to the intersection between these sets of macro-elements we can notice that \emph{resource} is the one that shares most of its elements with other sets (\emph{actor} and \emph{\grtext{data object}} in particular). This happens because elements such as \texttt{information} or \texttt{process participant} can indeed play different roles in a business process, acting e.g., as an artifact (resp., actor) or as a resource. 

Focusing on single
 elements, we can notice the big presence of \texttt{activity}, and the fact that most of the meta-models present a distinction between atomic and compound activities. \texttt{actor} and \texttt{role} are two other recurring elements together with \texttt{organisation}, if we sum it up also with \texttt{organisation unit}. \texttt{event} is another recurring element, together with \texttt{resource} and (data) \texttt{object}. The final group of recurrent elements is given by flow elements, where we can notice several flow and gateway elements appearing in at least \grtext{9} to \grtext{16} meta-models. Again, very few meta-models mention \texttt{goal} and \grtext{value-related} elements. 

Finally, only 14 elements appear in \grtext{at least the 25\%} of the primary studies. They are denoted in bold in Table~\ref{elementsRQ2} and are: \texttt{activity}, \texttt{atomic activity},
 \texttt{compound activity}, \texttt{event-EPC}, \texttt{event-BPMN}, \texttt{precondition},
 \texttt{gateway}, \texttt{parallel gateway}, \texttt{exclusive gateway}, \texttt{inclusive gateway}, \texttt{artifact}, 
 \texttt{actor}, \texttt{role}, and \texttt{resource}. Only 1 element (\texttt{activity}) appears in more than 50\% of the studies. 

\subsection{Answering \rqthree}
\label{subsec:RQ3}

The aim of this research question
 is to identify the reason to introduce/use the meta-models in the selected primary studies. Note that the reason to introduce the meta-model does not necessarily coincide with the overall aim of the paper. In fact, the meta-model is often an instrument for reaching a more comprehensive goal rather than being the goal of the paper.   

Table~\ref{table:role} provides a categorisation of the primary studies w.r.t. \grtext{17} 
 different purposes we were able to extract from the studies themselves. While extracting the reason to introduce a meta-model is somehow complex, as meta-models can be exploited in several ways, in the table we report only the purposes that were actually substantiated and illustrated in the papers, and not, for instance, to the ones that were just mentioned or left for future work and generalisations. 

\begin{table}[ht]
\centering
\scalebox{.8}{
\begin{tabular}{ll}
\toprule
  Class & Primary studies  \\ \midrule    
	\textbf{\grtext{describe what a business process is}} & \grtext{all}\\
  \textbf{extend a meta-model with new concepts} & \cite{DBLP:conf/caise/HeidariLK10,DBLP:journals/ijbpim/RosemannRF08,DBLP:conf/iceis/BouneffaA13,DBLP:conf/isda/HassenTG16,martins2015business,DBLP:conf/caise/WeigandJABEI06,DBLP:journals/jkm/PapavassiliouM03,DBLP:conf/caise/RussellAHE05,DBLP:journals/infsof/ArevaloCRD16,hassen2017extending,DBLP:conf/wi/DorndorferS17,DBLP:conf/iceis/KorherrL07,DBLP:conf/er/RosaDHMG08,DBLP:conf/es/SprovieriV15,DBLP:journals/infsof/StrembeckM11,DBLP:conf/caise/RuizCEFP14}\\
  \textbf{incorporate patterns in meta-model} & \cite{thom2005improving,krumeich2016modeling,DBLP:journals/infsof/StrembeckM11}\\
  \textbf{integrate process \& domain ontology} & \cite{DBLP:conf/icis/HuaZS10,DBLP:journals/jodsn/CherfiAC13}\\
  \textbf{support quality of models} & \cite{DBLP:conf/caise/HeidariLK10,DBLP:conf/icis/HuaZS10,DBLP:journals/jodsn/CherfiAC13,thom2005improving}\\
  \textbf{compare modelling languages} & \cite{DBLP:conf/caise/SoderstromAJPW02, DBLP:conf/wecwis/HeidariLBB13, DBLP:conf/sac/ListK06,DBLP:conf/wise/KunchalaYY14}\\
  \textbf{map/integrate modelling languages} & \cite{DBLP:conf/wecwis/HeidariLBB13,DBLP:conf/bpm/GrangelCSP05}\\
  \textbf{classify modelling languages} & \cite{DBLP:conf/caise/SoderstromAJPW02}\\
	\textbf{evaluate modelling languages} & \cite{DBLP:conf/sac/ListK06,DBLP:conf/wise/KunchalaYY14}\\
	\textbf{create language independent representation} & \cite{DBLP:conf/wecwis/HeidariLBB13,DBLP:journals/ijbpim/AxenathKR07,DBLP:conf/edoc/BruningG11,DBLP:conf/ecmdafa/HolmesTZD08}\\
  \textbf{describe a modelling language} & \cite{DBLP:journals/is/DamaggioHV13, Farrell2006FormalisingW}\\
  \textbf{define a new modelling language} & \cite{DBLP:conf/adbis/MomotkoS04,DBLP:journals/scp/MosserB13}\\  
	\textbf{clarify semantics of modelling language} & \cite{DBLP:conf/sac/SantosAG10}\\	
	\textbf{formal representation} & \cite{DBLP:conf/bpmn/Natschlager11,DBLP:conf/dexa/NicolaMPS10}\\
  \textbf{exploit automated reasoning} & \cite{DBLP:conf/bpmn/Natschlager11,DBLP:conf/dexa/NicolaMPS10}\\
  \textbf{evaluate suitability of a ML for a domain} & \cite{DBLP:conf/hicss/WeissW11a}\\
  \textbf{support extension of a ML to a new domain} & \cite{DBLP:conf/hicss/WeissW11a}\\
 \bottomrule
\end{tabular}}
\caption{Why introducing meta-models?}
\label{table:role}
\end{table}

As we can see, all meta-models in our primary studies aim at providing an illustration of what a business process is. The second 
 most popular usage of a meta-model in our primary studies was the extension of the meta-model itself with a new concept (16 papers). \cite{DBLP:conf/caise/HeidariLK10} extends it with quality metrics; \cite{DBLP:journals/ijbpim/RosemannRF08,DBLP:conf/wi/DorndorferS17} with a notion of context; \cite{DBLP:conf/iceis/BouneffaA13} with the notion of change and how change relates to business process elements; \cite{DBLP:conf/isda/HassenTG16,DBLP:journals/jkm/PapavassiliouM03,hassen2017extending} with the notion of knowledge and knowedge-related concepts; \cite{martins2015business} introduces the relation between business processes and daily practices; \cite{DBLP:conf/caise/WeigandJABEI06} extends a business process meta-model with the notion of value; \cite{DBLP:conf/caise/RussellAHE05,DBLP:conf/er/RosaDHMG08} with the notion of resource; \cite{DBLP:conf/er/RosaDHMG08} introduces also a data dimension concerning artefacts and data objects; \cite{DBLP:journals/infsof/ArevaloCRD16} with the notion of time; \cite{DBLP:conf/es/SprovieriV15,DBLP:conf/iceis/KorherrL07} extends it with the notion of goal, and \cite{DBLP:conf/iceis/KorherrL07} enriches it also with the notion of performance; finally, \cite{DBLP:journals/infsof/StrembeckM11} extends it with RBAC related concepts (e.g., roles) and also RBAC related workflow patterns. 
Examples of extension of the meta-model are even more present if we consider also the two additional papers that incorporate workflow patterns in the meta-model and the \grtext{two} papers that extend business process meta-models with the ability to connect to domain ontologies.  

Coming to the less frequent usages we can note that 7 papers exploit meta-models for comparing (integrating, classifying) different modelling languages and in some cases evaluate them; instead, 8 papers use meta-models for describing an existing modelling language, support the definition of a new one, or create from them a language independent representation. 
 Another group of papers (\grtext{3} in total) focuses on the creation of formal representations of meta-models in order to clarify the semantics of specific modelling languages or exploit automated reasoning techniques (e.g., to verify the well formedness of a business model specification). 
 One paper exploits the business process meta-model of the Semantic BP Modeling Language (tipically used to model the public sector domain) to evaluate its adequacy to the banking sector, and to find out requirements for the modification of the language to the new domain. 

\subsection{Answering \rqfour}
\label{subsec:RQ4}

As already reported in Table~\ref{table:Addtable}, few primary studies present some forms of evaluation of the meta-models they describe. In answering \rqfour we aim at investigating the way these evaluations are carried out. 

Table~\ref{table:evaluation} provides a categorisation of the forms of evaluation we were able to extract from the primary studies.  Given that not many papers provide in depth evaluations, we have listed here also the studies in which use cases are mainly used as illustrative examples of how the meta-model (or the framework that includes the meta-model) can be applied.

\begin{table}[ht]
\centering
\scalebox{.8}{
\begin{tabular}{ll}
\toprule
  Class & Primary studies  \\ \midrule    
  \textbf{Extensive Case Studies} & \cite{DBLP:conf/isda/HassenTG16,DBLP:journals/scp/MosserB13}\\
  \textbf{Ontological Analysis} & \cite{DBLP:conf/wecwis/HeidariLBB13,DBLP:conf/sac/SantosAG10}\\
  \textbf{Comparison with requirements} & \cite{DBLP:conf/wi/DorndorferS17,DBLP:conf/hicss/WeissW11a}\\
  \textbf{Formal properties} & \cite{DBLP:conf/bpmn/Natschlager11} \\
	\textbf{Illustrative examples} & \cite{DBLP:conf/wecwis/HeidariLBB13,DBLP:conf/icis/HuaZS10,DBLP:journals/jodsn/CherfiAC13,DBLP:journals/ijbpim/RosemannRF08,DBLP:conf/edoc/BruningG11,DBLP:conf/caise/WeigandJABEI06,DBLP:journals/jkm/PapavassiliouM03,DBLP:journals/infsof/ArevaloCRD16,DBLP:conf/iceis/KorherrL07,krumeich2016modeling,DBLP:journals/infsof/StrembeckM11,DBLP:conf/bpm/GrangelCSP05}\\
 \bottomrule
\end{tabular}}
\caption{How are the meta-models evaluated?}
\label{table:evaluation}
\end{table}

Overall, only \grtext{7} papers present some form of evaluation, while 12 papers present illustrative examples.
Illustrative examples are, thus, the most recurring method to show the applicability of the approach.  
\cite{DBLP:conf/wecwis/HeidariLBB13} provides a demonstration of applicability of the business process ontology it introduces to represent business process models by using a \textit{Processing of automobile insurance claim}
 example. \cite{DBLP:conf/icis/HuaZS10} provides an illustration of how an online auction process is modelled using the approach presented in the paper. This illustration concerns also the meta-model as it shows how the ontology for the use case is built using the meta-model.  \cite{DBLP:journals/jodsn/CherfiAC13} exploits a use case to illustrate both the alignment between the  domain ontology and the business process model and the fact that incorporating a domain ontology improves the quality of the resulting models. \cite{DBLP:journals/ijbpim/RosemannRF08} provides a case study which illustrates how the framework can be applied to model a ticket reservation and check-in process of a major Australian airline, and in particular to model the contextual dependent aspects of this process. \cite{DBLP:conf/edoc/BruningG11} presents a use case in the medical domain to illustrate how the meta-model is used to model an actual workflow with respect to data and organisational aspects. \cite{DBLP:conf/caise/WeigandJABEI06} provides a use case taken from a scientific conference scenario to illustrate how the value object model presented in the paper can support the production of a Value Resource Model for the specific use case. \cite{DBLP:journals/jkm/PapavassiliouM03} provides an illustration of how the modelling tool based on the theoretical meta-model proposed in the paper can be used to model the granting of full old age pension within the Greek Social Security Institute. In \cite{DBLP:journals/infsof/ArevaloCRD16} a short illustration of how the motivating example is modelled using the proposed framework (based on the extended meta-model with time constraints) is presented. \cite{DBLP:conf/iceis/KorherrL07} demonstrates the practical applicability of the extension of the extended EPC and BPMN meta-models with an application to the \emph{Processing of Automobile Claims} business process; \cite{krumeich2016modeling} introduces an example to show how the modelling technique based on the extended metamodel of EPC to represent complex events can be used to represent an exemplary complex event pattern. \cite{DBLP:journals/infsof/StrembeckM11} provides several real case examples to discuss how the newly introduced concept of Business Activity can be used to define process-related RBAC models. Finally, \cite{DBLP:conf/bpm/GrangelCSP05} presents a use case to validate the POP* meta-model as a common and standard language to exchange models among different Enterprise Modelling Tools.

The only two papers that provide real/extensive use cases and exploit them to support precise characteristics of the meta-model based framework are \cite{DBLP:conf/isda/HassenTG16} and \cite{DBLP:journals/scp/MosserB13}. The first presents a real use case taken from a medical domain. Here the aim is to go beyond a mere illustration and to evaluate how the concepts contained in the meta-model can support an understandable, adequate and expressive representation of Sensitive Business Processes. The latter provides an extensive validation of the Adore method (including the Adore meta-model) against two large use cases with the aim of showing that it can be used in a real-life context, and that it supports the capture of a real-life evolution process at the business process level.

A different form of evaluation of the characteristics and quality of the meta-models is provided in \cite{DBLP:conf/wecwis/HeidariLBB13} and \cite{DBLP:conf/sac/SantosAG10}. These primary studies  exploit an ontological analysis to show how the meta-meta model is successful in expressing concepts taken from upper level ontologies. In the first paper the upper level ontology used is the Bunge-Wand-Weber (BWW) upper level ontology~\cite{Wand:1990ab}, while in the second it is the UFO upper-level ontology~\cite{Guizzardi-Wagner-UFO}. 

\cite{DBLP:conf/wi/DorndorferS17} provides an evaluation of the extended meta-model by comparing it with the requirements for its development presented at the beginning of the paper. A similar evaluation is provided in \cite{DBLP:conf/hicss/WeissW11a}. 

Finally, \cite{DBLP:conf/bpmn/Natschlager11} provides an evaluation of the formal ontology in terms of its formal (logic-based) properties of consistency and correctness. 

By looking at these results we can say that a rigorous evaluation of meta-models is often neglected in literature as it reduces, in the majority of cases, to mere illustrative examples. 
Three forms of evaluation stand out from this analysis and can provide the basis for 
guidelines and evaluation criteria for the development of meta-models in the area of business processes. First, an evaluation by means of real use cases: this can help the assessment of the elements contained in the meta-model to support the modeling of real scenarios. Second, an evaluation by means of a comparison with requirements: this can help the assessment of the meta-model w.r.t. needs or conditions that motivated its development. Third, an evaluation based on foundational ontologies: this can help assessing the meaning and properties of concepts present in the meta-model on the basis of well-known reference elements contained in foundational ontologies.

\section{Discussion of Results} 
\label{sec:discussion}

The data presented in Section~\ref{sec:RRQs} enable answering, at least partially, the four research questions presented in Section~\ref{sec:RQ} that were used to shape this SRL. 

Before addressing the research questions in detail, let us comment on the temporal distribution and the distribution by publication type of the primary studies.
Concerning the temporal distribution we can observe that, while we did not pose any temporal restriction towards the data search in Scopus, WoS, and DBLP, and while we also manually evaluated all the CAiSE proceedings from 1990, the first paper included in our primary studies was published quite recently in 2002. It also takes until 2005 to have more than 1 paper per year included into the set. Thus, the interest in this area seems to be a recent one\footnote{Interestingly enough, 2003 was the year when the BPM conference series started.} with slightly more than 50\% of works published in 6 years between 2007 and 2013. 
Overall, the relatively low number of papers identified, and their temporal distribution indicate that this topic is still under-investigated. Also, considering the importance of the topic, and the growing interest in different approaches towards Business Process Modelling (e.g., procedural vs.~declarative or activity-centric vs.~artefact-centric styles of modelling) we were expecting a larger number of publications in the last 10-year period, with a growing trend. Instead, we notice a slight decrease of publications in the last \grtext{four} 
 years, which could be related to this lack of a comprehensive common ground where to place new proposals of meta-models.  

Regarding the distribution by publication type, we can notice a reasonable indication of scientific maturity. Indeed, the data contained in \ref{app:sourcelist} show a good distribution between journal and conference publications (\grtext{28\%} and \grtext{58\%} of the total, respectively), and - even more important - \grtext{1/4}
 of \grtext{the} primary studies (\grtext{25\%}) that was published in journals/conferences ranked Q1 or A/A$^*$ according to Scopus/CORE (see details in~\ref{app:sourcelist}). This number increases to \grtext{about 42\%} if we include also journals/conferences ranked Q2 or B. 
If we restrict only to journal publications, \grtext{6} out of \grtext{10} (\grtext{60\% of the total number of journals}) belong to the 1st or 2nd Quartile according to the chosen journal ranking. 
Not surprisingly, the publication venues 
 mostly refer to the areas of Software Engineering, Conceptual Modelling, and Business Process Management, even though no standard venue was identified as a target for the authors of such primary studies. A notable exception are the 4 papers published in the CAiSE (International Conference on Advanced Information Systems Engineering) conference, which represent \grtext{11\%} of the total, while, surprisingly, no papers were published in the BPM (International Conference on Business Process Management) conference. 

Concerning the research questions, which are the targets of this SLR, the numeric results and some comments are already contained in Section~\ref{sec:RRQs}. We report here some additional remarks that mainly highlight the overall findings and the limitations of current published research.

Focusing on \rqone, it is interesting to notice that \grtext{half} of the primary studies do not target any specific business process modelling language. This means that the description of what constitutes a business process is perceived as a topic of research \emph{per se}, and is not necessarily tight to a specific modelling language or approach. 
Also, most of the primary studies that
 focus on specific modelling languages target existing languages. This seems to indicate a reasonable maturity and level of satisfaction towards the available modelling languages. 
By looking into the characteristics investigated in Table~\ref{table:Addtable} at page \pageref{table:Addtable} we can note that \grtext{83\%} of the primary studies consider (at least) the process model dimension and \grtext{50\%} consider (at least) the execution dimension, with \grtext{33\%} considering both. While the first result is perhaps not very surprising, we consider very positive the conspicuous presence of studies that incorporate also the execution dimension. Indeed, executions of processes are, in the BPM fields, regarded as first class citizens and not simply as mere instances of process models. As an example consider the importance of process executions (a.k.a
 event logs) in the field of Process Mining.
Another interesting result is the one that refers to the approaches taken towards business process modelling. As already said in Section~\ref{subsec:RQ1} 
most primary studies adhere to the traditional procedural and activity-centric based view on business processes with very few papers taking a declarative or artefact-centric view (\grtext{19\%} and 8\% of primary studies, respectively). Concerning the lack of meta-models of declarative languages / models, the result becomes even more interesting if we consider that no primary study is devoted to the investigation of meta-models of the DECLARE modelling language, despite its 
 growing popularity in the scientific BPM community.   
Further interesting data concern the domain (in)dependency of meta-models. Indeed only \grtext{about 5\%} of the primary studies address domain dependent business process models. Thus, we can say that an effort to describe what constitutes a generic business process is well under way. Instead, investigations of what constitute a business process in a specific domain (e.g., an administrative procedure, a retail oriented business process, just to mention two popular domains) is way less clear and investigated.  

Focusing on \rqtwo, a detailed analysis is already reported in Section~\ref{subsec:RQ2}. Summarising, the results shown in Table~\ref{elementsRQ2} at page \pageref{elementsRQ2}
indicate that the elements of the process control flow (\emph{activity} and \emph{sequence flow}) together with the \emph{data object} and the organisational dimension (\emph{actor}) are the most recurring both in terms of overall presence and decomposition into different elements. Instead \emph{goal} and \emph{value} aspects are poorly \grtext{and, in case of \emph{value}, even not explicitly} described both in terms of occurrences in primary studies meta-models and decomposition.
 This finding is certainly correlated to the fact that most business process modelling languages do not include values and/or goals in the graphical design of a business process model. Nonetheless, it is easy to observe that the situation does not change if we consider language-independent meta-models. This somehow clashes with most of the modern definitions of Business Process, which explicitly mention either the (added) value brought by the process execution\footnote{As an example, Johansson et al. \cite{Johansson:1993aa} says that a business process is ``''\emph{a set of linked activities that take an input and transform it to create an output. Ideally, the transformation that occurs in the process should add value to the input}''.} or the goal a process execution has to realise\footnote{See e.g., the definition taken from \cite{DBLP:books/daglib/0029914} and reported in Section~\ref{introduction} at page \pageref{def:process}.} 
Thus, while it seems to be ``extremely clear and well agreed that business processes realise a business goal", as recently highlighted in~\cite{AdamoAIIA2018}, it appears to be more difficult to leverage state-of-the-art business process meta-models to state exactly what this business goal (resp. value) is and which characteristics it detains, as recently highlighted in~\cite{AdamoAIIA2018}. 

If we focus on the elements appearing in at least 25\% of the studies, we can notice a high presence of elements related to the control flow w.r.t. other aspects of the business process. On the positive side, the elements related to the control flow that appear at least in 25\% of the studies\footnote{The control flow elements that appear in at least 25\% of the studies are \texttt{activity}, \texttt{atomic activity}, \texttt{compound activity}, \texttt{event-EPC}, \texttt{event-BPMN}, \texttt{gateway}, \texttt{parallel gateway}, \texttt{inclusive gateway} and \texttt{exclusive gateway}.} correspond to key elements of a business process control flow. On the negative side, in addition to the \texttt{goal} (\texttt{value}) aspect already discussed above, it is interesting to notice that while \texttt{actor} and \texttt{role} are present in \grtext{25\%}  of the studies, \texttt{organization} is not, and  \texttt{artifact} is the only data object element with more that 25\% of presence. If we instead consider the elements appearing in at least 50\% of the studies, which could be considered a ``core'' set of elements of what constitutes a business process, we reduce to only \texttt{activity}. In our opinion this is a sign of a lack of a mature answer to the fundamental question of ``what constitutes a business process'' and an evidence of the fact that most works have mainly addressed business processes just looking at control flow related aspects, somehow neglecting a comprehensive investigation which takes into account all the characterising aspects of this notion. 

The results of \rqthree, summarised in Table \ref{table:role}, show an interesting and articulated usage of business process meta-models. Even though the most popular usage of meta-models is somehow self referential (``extension of the meta-model with a new concept''), the number of other usages denote a reasonable maturity in the field, in particular for what concern the exploitation of meta-models to investigate aspects of specific business process modelling languages. A possible limitation here is the lack of foundational studies that address the fundamental question of what a business process is and what differentiates it from other kinds of processes. 

The results of \rqfour, summarised in Table \ref{table:evaluation}, are on the problematic side. Indeed only \grtext{less than 6\%} of \grtext{the} primary studies show an extensive evaluation phase with real case studies, and another \grtext{11\%} show an evaluation of the adequacy of the meta-model using reference foundational ontologies or a comparison with requirements. This lack of coverage of the ``evaluation'' phase may be justified by many different factors: on the one hand evaluating the adequacy, or usefulness, of a generic meta-model in concrete domain-specific scenarios is a complex activity, especially when there are no standard reference scenarios for this activity; on the other hand meta-models are introduced for different purposes (see the answer to \rqthree) and different purposes may require different evaluation strategies. 
 This finding highlights a limitation of current research, and the BPM community should make an effort to understand whether (and how) an evaluation of meta-models could be carried out. Nonetheless, the three typologies of evaluation present in the primary studies (evaluation with real case studies; evaluation of the adequacy using reference foundational ontologies; and comparison with requirements) provide a good starting point for this discussion. 

\subsection{Limitations of this study} 
\label{sub:limitations_of_this_study}

The main limitations of this study are common to the literature reviews and include (i) biases in the selection of the papers; (ii) imprecisions introduced in the extraction of data from the selected works; (iii) potential inaccuracies due to the subjectivity of the analysis carried out. 

To mitigate these threats, we followed the guidelines reported in~\cite{Kitchenham07guidelinesfor,kitchenham2004procedures}. 
We applied the standard procedures reported in the guidelines for the correctness of the SLRs, such as the identification of the proper keywords to perform the data search, the selection of the appropriate sources and repositories for the field under investigation, the definition of clear inclusion and exclusion criteria, as well as of the quality assessment parameters.
Specifically, we relied on the main literature sources and libraries in the information system field for the extraction of the works related to business process models and meta-models. Moreover, we expanded the search by manually inspecting the two main reference conferences in the field of BPM. To further improve the reliability of the review, we put some effort in guaranteeing the reproducibility of the search by other researchers, although ranking algorithms used by the source libraries could be updated and provide different results. 

A further limitation of this study lies in the facts that: (i) only one researcher selected the candidate primary studies;
 and furthermore (ii) only one researcher worked on the data extraction. Nevertheless, both aspects have been mitigated by the fact that (i) another researcher checked the inclusion and the exclusion of the studies; and (ii) another researcher checked the data extraction, as suggested in~\cite{BRERETON2007571}.

\section{Conclusions}
\label{sec:conclusion}

This work provides the first systematic literature review of business process meta-models. This systematic literature review addressed research questions concerning (i) the kind of meta-models proposed in literature; (ii) the recurring constructs they contain; (iii) their purpose(s); and (iv) their evaluations. 

The analysis provided in this SLR shows that there is a reasonable body of work conducted in this specific area, even though the field does not appear to have reached full maturity. On the positive side, a reasonable number of high quality publications exist in literature, which present well described business process meta-models. These meta-models are almost equally targeting specific BPMLs or the notion of business process in general. Also, they cover both the model and execution aspects of business processes. Another positive aspects are the number of different reasons for introducing/exploiting these meta-models, which is an evidence of liveliness of the topic, and the reasonable presence of key control flow elements in the meta-models. Also, some good examples of how to evaluate meta-models are present in literature.   
On the negative side we can notice: a lack of meta-models for the ``new'' paradigms
 towards business process modelling, namely, the declarative based and artefact centric approaches; a lack of presence of non control flow key aspects of business processes in meta-models; and a lack of evaluation of meta-models in literature.   
These results could open up an opportunity for new research efforts addressing these aspects.

The analysis provided in this SLR could be used as a starting point to define a framework for the description and classification of business process meta-models. Indeed, the characteristics identified in answering \rqone, \rqthree, and \rqfour provide an extensive set of ``tags'' which could be used to annotate meta-models, while analogous ``tags'' to describe the content could be defined starting from the answer to \rqtwo . These annotations could be, in turn, used to retrieve meta-models with specific characteristics, or to compare and analyse them further in the future.
Similarly, the analysis of the meta-model elements produced in answering \rqtwo could be used as a starting point for defining an ``emerging'' business process meta-model from data. To do that, an analysis of the relationships between these elements (or at least between the most recurring ones) should be produced, and this is part of a work we would like to start in the immediate future.

\appendix
\section{Primary Studies' Publication Venues}
\label{app:sourcelist}
\begin{center}
\scriptsize
\begin{longtable}{p{10cm} p{1cm} }
	\toprule
	Journal & Paper \\ 
	\midrule
	Information and Software Technology$^{**}$ & \cite{DBLP:journals/infsof/ArevaloCRD16,
	DBLP:journals/infsof/StrembeckM11}\\
	International Journal of Business Process Integration and Management & 	\cite{DBLP:journals/ijbpim/AxenathKR07, DBLP:journals/ijbpim/RosemannRF08} \\
	Journal on Data Semantics$^{*}$ & \cite{DBLP:journals/jodsn/CherfiAC13} \\
 	Journal of Knowledge Management & \cite{DBLP:journals/jkm/PapavassiliouM03}\\
 	Information Systems$^{**}$ & \cite{DBLP:journals/is/DamaggioHV13} \\
	Science of Computer Programming & \cite{DBLP:journals/scp/MosserB13}  \\ 
	Procedia Computer Science$^{*}$ & \cite{martins2015business} \\
	Group Decision and Negotiation$^{*}$ & \cite{Farrell2006FormalisingW} \\
	\toprule
	Conference \& Symposium & Paper \\ 
	\midrule
	International Conference on Business Informatics & \cite{DBLP:conf/wecwis/HeidariLBB13} \\
	International Conference on Intelligent Systems Design and Applications & \cite{DBLP:conf/isda/HassenTG16} \\
	International Conference on Database and Expert Systems Applications$^{*}$ &	\cite{DBLP:conf/dexa/NicolaMPS10} \\
	International Conference on Conceptual Modeling$^{**}$ & \cite{DBLP:conf/er/RosaDHMG08} \\
	International Conference on Information Systems$^{**}$ & \cite{DBLP:conf/icis/HuaZS10} \\
	East European Conference on Advances in Databases and Information Systems & 	\cite{DBLP:conf/adbis/MomotkoS04} \\
	International Conference on Enterprise Information Systems & 	\cite{DBLP:conf/iceis/KorherrL07,DBLP:conf/iceis/BouneffaA13} \\
	International Conference on Advanced Information Systems Engineering$^{**}$ & 	\cite{DBLP:conf/caise/SoderstromAJPW02, DBLP:conf/caise/RuizCEFP14, 	DBLP:conf/caise/RussellAHE05, DBLP:conf/caise/WeigandJABEI06} \\
	European Conference on Model Driven Architecture-Foundations and Applications & 	\cite{DBLP:conf/ecmdafa/HolmesTZD08} \\
	International Conference on Enterprise Systems  & \cite{DBLP:conf/es/SprovieriV15} \\
	Hawaii International Conference on System Sciences & \cite{DBLP:conf/hicss/WeissW11a} \\
	International Enterprise Distributed Object Computing Conference$^{*}$ & 	\cite{DBLP:conf/edoc/BruningG11} \\
	Multikonferenz Wirtschaftsinformatik & \cite{krumeich2016modeling}\\
	Internationale Tagung Wirtschaftsinformatik & \cite{DBLP:conf/wi/DorndorferS17} \\
	ACM Symposium on Applied Computing$^{*}$ & \cite{DBLP:conf/sac/ListK06, 	DBLP:conf/sac/SantosAG10}\\
	International Symposium on Business Modeling and Software Design & 	\cite{hassen2017extending} \\
	\toprule
	Workshop  & Paper \\ 
	\midrule	
	International Workshop on Personalization and Context-Awareness in Cloud and Service Computing & 	\cite{DBLP:conf/wise/KunchalaYY14}\\
	Workshop on Enterprise and Organizational Modeling and Simulation & 	\cite{DBLP:conf/caise/HeidariLK10} \\
	Workshop on Business Process Intelligence & 	\cite{DBLP:conf/bpm/GrangelCSP05} \\
	Workshop XML for Business Process Management & \cite{thom2005improving} \\
	International Workshop on Business Process Modeling Notation & 	\cite{DBLP:conf/bpmn/Natschlager11} \\
	\bottomrule
\end{longtable}
\end{center}

The venues marked with $^{**}$ are classified as Quartile 1 (Q1) or A/A$^*$ according to the Scopus journal ranking 2017 and the CORE conference ranking 2017, respectively.   
The venues marked with $^{*}$ are classified as Quartile 2 (Q2) or B according to the Scopus journal ranking 2017 and the CORE conference ranking 2017, respectively.

\bibliographystyle{spbasic}

\end{document}